\documentclass[12pt]{iopart}

\usepackage{graphicx}
\usepackage{bm}
\usepackage{color}

\newcommand{\be}{\begin{equation}}
\newcommand{\ee}{\end{equation}}
\newcommand{\ba}{\begin{eqnarray}}
\newcommand{\ea}{\end{eqnarray}}
\newcommand{\bd}{\begin{displaymath}}
\newcommand{\ed}{\end{displaymath}}
\renewcommand{\vec}[1]{\mbox{\boldmath$#1$}}

\begin{document}

\title[Collective Flow]{Global Collective Flow in Heavy Ion Reactions\\
from the Beginnings to the Future}

\author{L.P. Csernai$^1$ and H. St\"ocker$^{2,3}$}

\address{
$^1$ Department of Physics and Technology, University of Bergen,
Allegaten 55, 5007 Bergen, Norway\\
$^2$ Gesellschaft f\"ur Schwehrionenforschung, Planckstr. 1, Darmstadt,
Germany\\
$^3$Frankfurt Institute for Advanced Studies - Goethe University, 
60438 Frankfurt am Main, Germany
}
\ead{csernai@ift.uib.no}

\begin{abstract}
Fluid dynamical models preceded the first heavy ion accelerator
experiments, and led to the main trend of this research since then.
In recent years fluid dynamical processes became a dominant direction of
research in high energy heavy ion reactions. The Quark-gluon Plasma formed
in these reactions has low viscosity, which leads to significant fluctuations
and turbulent instabilities. One has to study and separate
these two effects, but this is not done yet in a systematic way.
Here we present a few selected points of the early developments,
the most interesting collective flow
instabilities, their origins, their possible ways of detection and 
separation form random fluctuations arising from 
different origins, among these the most studied is the randomness of 
the initial configuration in the transverse plane.
\end{abstract}


\maketitle
%
\section{Introduction}
\label{intro}

The  
acceleration of heavy nuclei to relativistic energies started in the
1970s. At that time the expectation was that the nuclear matter
may be compressed and heated up in such collisions to states known to
exist in neutron stars, in back holes or in the early universe only.
The majority of the nuclear physics community was rather skeptical about these
expectations, many physicists claimed that heavy ion collisions will
not provide new information and only a straightforward multiplication
of the well known nucleon - nucleon collisions will occur. Only a few
theorists and experimentalists advocated persistently the idea that
collective effects, the consequences of the large collective pressure,
shock waves and collective flow should be observable in heavy ion reactions.

The unique role of fluid dynamics (FD) approach was obvious from the 
beginning. This is the only model, which  hast the advantage that it 
uses the Equation of State (EoS)
as direct input characterizing the matter, and provides
the direct consequences of these matter properties. This advantage is
lost with some modifications: as anisotropic fluid dynamics or two- or
more fluid dynamical models, which cannot make use of a unique EoS.

It took more than 10 years, the construction of several revolutionary new
detectors, the development of many new evaluation techniques until in
1984 the existence of the collective flow was proven beyond any doubt.
This news made it even to the New York Times, and later this discovery
was honored by several national prizes. 

By now these collective flow effects are widely used to extract the 
compressibility and viscosity of nuclear matter from collision measurements.
This success served as basis of the present progress where we search
for the phase transition into the Quark - Gluon Plasma at CERN and BNL.
At both laboratories there are large scale experimental investments into
relativistic heavy ion physics, and by now thousands of researchers are
working intensively on this field worldwide. The pioneers of this field
certainly deserve the highest recognition for their devoted and persistent
work and success.

\subsection{Early Developments}
\label{early}

If one has to select the main figures of the field the first and
most persistent theoretical predictions came from Walter Greiner's 
institute. He was author of the very first publications himself. The most
important of these early works is W. Scheid, H. M\"uller, and W. Greiner's
paper in 1974 in Phys. Rev. Lett. \cite{g1}.
Here the compression of nuclear matter
in shock waves was first suggested as a basic process in 
heavy ion collisions.
There were of course many other
publications related to this subject both before and especially after this 
work. The preceding works by Scheid, Ligensa and Greiner discussed nuclear 
compressibility in such collisions, while in another early subsequent work
of H. Baumgardt, {\it et al.} \cite{g2} possible experimental 
consequences were already discussed.

At the same time the possibility of shock waves in an independent work by
G.F. Chapline, M.H. Johnson, E. Teller, and M.S. Weiss,
\cite{g3}, was pointed out also. However, this group did not follow the further
development so closely later and did not help the experimental effort
to the same extent as Greiner's group. Greiner and his coworkers worked out
numerous details in a very large number of works.

In these early years the experimental developments are also 
very important. Several competing groups worked on the problem.
The first real unambiguous result was achieved by Prof.  Hans Gutbrod's
group: H. \AA . Gustafsson, {\it et al.}, \cite{g4}. 
This group constructed a large new detector in Berkeley ``The Plastic
Ball'' and this helped them to achieve undebatable results. Other
leading scientists in the group were A. Poskanzer and H.G. Ritter.
By the way the group, as WA80, worked still together at CERN (including
the Plastic Ball) and it had a large contribution of more than
20 physicists. Soon after other groups
with other type of equipment confirmed the results, and as we mentioned in
the beginning, by now this phenomenon is used for practical purposes
to study the details of the equation of state of nuclear matter.

Although here we concentrate on the flow dynamics in this work
we have to mention that in the early decades of heavy ion physics
at the MSU and Bevelac energies important advances were reached in 
the nuclear matter EoS and its parameters, for example, the 
compressibility coefficient (K). Later on at high energy
collisions we dealt with hadron resonance gas EoS and transition 
to QGP EoS etc. These were distinct physics developments over a couple
of decades, a substantial part of these developments is discussed
in ref. \cite{Cs94}.

The authors were also among the firsts working in this 
field. They were among
those who have focused on the collective flow in high energy collisions
in the last 30 years. During these years the importance of the 
collective flow became increasingly dominant.
The first evidence of the existence of the collective flow
was the bounce off or side splash effect, where the projectile and 
target matter deflected each other, from the original
beam ($z$- ) direction, into the transverse, ($x- $) direction
in the reaction plane, due to the large collective pressure in 
the shock-compressed overlap domain. Initially this effect
was measured by the average transverse momentum $<p_x (y)>$ of the
nucleons as a function of rapidity, $y$.

Later this flow component was called the "directed transverse
flow". As beam energies increased the overlap time 
and the overlap surface has decreased due
to the Lorentz contraction of the nuclei in the center of mass 
(CM) frame, so the momentum transfer in the
transverse direction decreased as $\gamma^{-2}$. 
At the same time the beam momentum increased, so
the directed flow in the transverse direction 
became less dominant. Due to the same reason the
other transverse flow effect, the
"squeeze-out" decreased as well.
On the other hand, in the transverse plane, 
the almond shape of the overlap spectator
region, in finite impact parameter collisions, 
became more significant. This shape led to a
dominant expansion in the direction of the 
largest pressure gradient, i.e. orthogonally to the
flat almond shape. At asymptotically large energies, 
and in the case of full transparency the
direction of expansion would fall exactly in 
the $\pm x$ direction, where $[x, z]$ is the reaction
plane. In fact such a dominant expansion was 
detected at CM rapidity and was called "elliptic flow".

\begin{figure*}[h]
\centering
\hskip -2cm
\includegraphics[width=9cm]{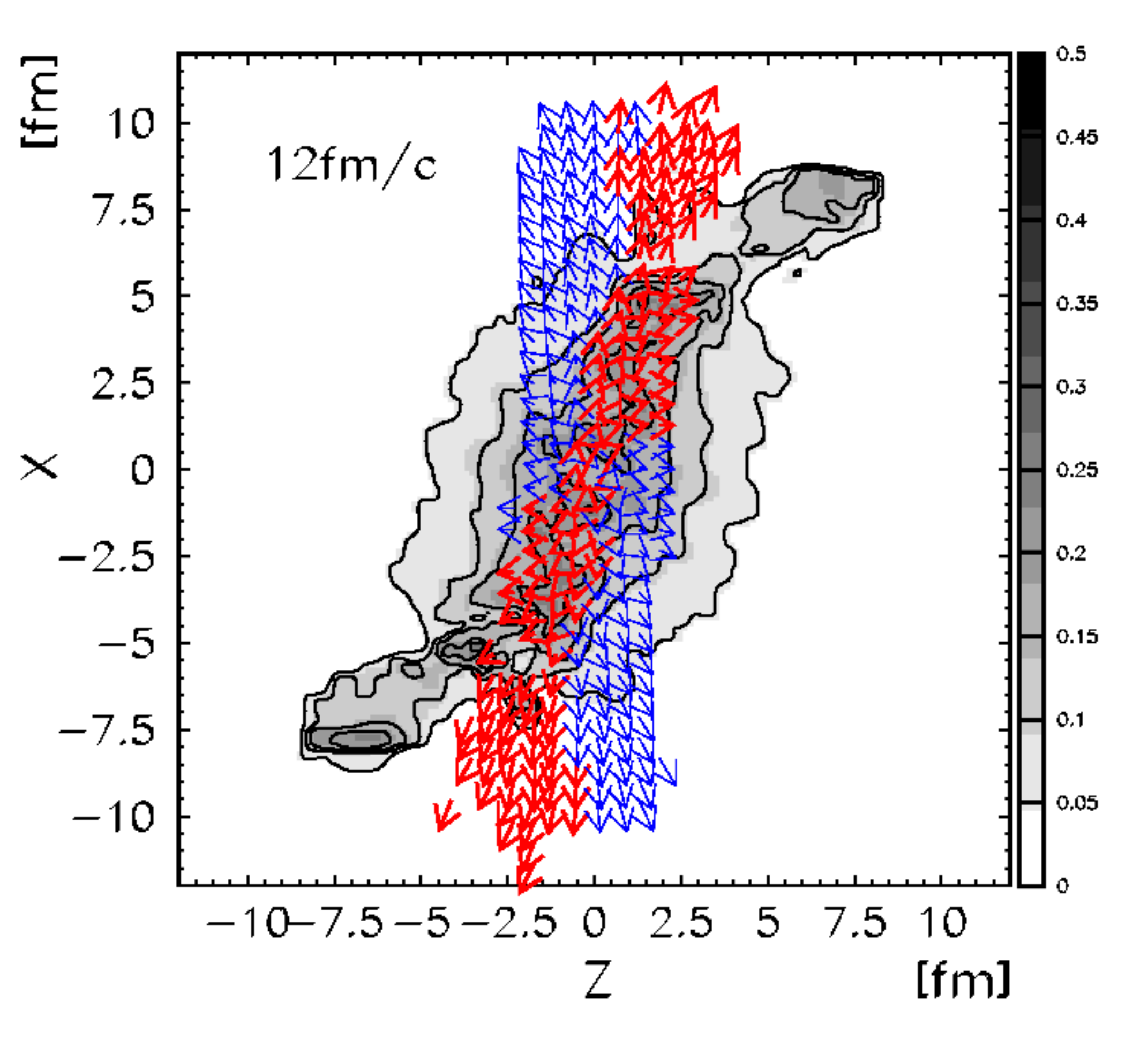}
\caption{
(color online)
Net-baryon density for a $Au+Au$  reaction at 8 A GeV, and $b=3$ fm,
with an EoS with phase transition at $t=$12 fm/c in the reaction plane with
velocity arrows for mid-rapidity ($y<0.5$) fluid elements. Antiflow
is indicated by thin, blue arrows, normal flow is indicated by bold
red arrows. From ref. \cite{antiflow}.
}
\label{fig-0}       
\end{figure*}

At finite energies the almond shape overlap 
region is not necessarily symmetric in
forward/backward direction, thus, the direction 
of the dominant expansion and largest pressure
gradient may not point exactly to the $\pm x$ 
direction. If the flat, elongated disc, representing the
initial state of the hot, compressed matter, 
is tilted, in the direction of the directed flow, 
i.e. in the flow angle $\Theta_F$ (with respect 
to the beam $(z)$ axis, then the direction of the 
dominant flow will not point to the CM $x$ direction 
but a little backward, tilted by  $\Theta_F$ into the $-z$ 
direction.
This effect was described first in 1999 \cite{cr99}, 
as the "third flow component" and measured in
several experiments at SPS. Later it was also called
as "antiflow" \cite{antiflow}, see Fig. \ref{fig-0}.
With increasing energies, 
as the $\Theta_F$ flow angle decreases and
becomes immeasurable in collider experiments, the 
deviation of dominant flow direction from
the CM $\pm x$ direction, became more and more difficult to measure.
Later a simplified kinetic explanation was also brought up, which did not
explicitly assumed a phase transition, but implicitly assumed a
collision mechanism of highly contracted projectile and target nuclei.
The predicted mechanism was supported by RQMD calculations also
\cite{SnellingsEA-00-08}.

The detailed and quantitative analysis of flow 
patterns is of vital importance, as these
provide the possibly most direct information 
about the pressure and about the space-time
configuration of matter.

The equations of perfect FD are just the local 
energy and momentum conservation laws, where
we assume that the energy-momentum tensor is 
given by the EoS, in a simple
covariant form. The underlying assumption is 
that the system is in local thermodynamical
equilibrium. Thus, the most important input 
is the EoS, while recent high energy collisions
in the QGP domain point at the importance 
of transport properties, especially of low viscosity.

In the late 70s and early 80s the applicability and relevance
of fluid dynamics were frequently questioned. The early suggestions
of Mach cone and Mach shock waves, as well as the  experiments
in the early 70s mentioned above did not uniformly convince 
the researchers in the field. More detailed and more precise
modeling work was needed with new measurable signatures of the 
collective flow. Apart of the EoS describing the properties of
matter in local thermal equilibrium, the importance of transport
properties, especially the viscosity and transport
processes was recognized very early 
also.\cite{csg10.,csg13.,csg23.}

The other question was what are the most typical, measurable
flow phenomena.  As most collisions happen at finite impact parameters,
in these reactions instead of symmetric Mach cones, a bounce off,
side splash or directed flow develops in a well defined azimuthal
and polar direction in the reaction plane according to the predictions
of the Frankfurt group.  
\cite{csg14.,csg20.,csg21.,csg24.}
This typical flow pattern was also obtained independently
in Los Alamos model calculations
\cite{csl1,csl2}.
Still it had to be shown that the fluid dynamical predictions
are significantly different from predictions of random
nuclear cascade, without collective flow effects. This was
also done in a concerted theoretical effort
\cite{csg17.}. The resulting directed flow or bounce off
effect could then be observed in different measurable quantities, 
including two particle and multi-particle correlations, 
among nucleons and nuclear fragments
\cite{csg14.,csg20.,csg21.,csg24.}.

As mentioned before, the first comprehensive experimental work
that fully confirmed the existence of collective flow and the
bounce effect was published in 1984.
\cite{g4}
Shortly after the streamer chamber measurements could also
confirm the existence of collective flow, using a sensitive evaluation
method developed by Danielewicz and Odyniecz \cite{do85}.
This method was so sensitive that it could confirm the existence
of collective flow even in emulsion experiments of a sample
of about 100 events and only a few hundred particles!
\cite{cf86}

In the next decade we learned the basic features of the EoS
of hot and compressed nuclear matter, its compressibility.
Flow measurements have contributed to the discovery of
nuclear liquid-gas phase transition: at the National Superconducting 
Cyclotron Laboratory at Michigan State University, at energies 
around 50 MeV/nucleon negative flow was observed as a consequence of 
the phase transition.

Even more interesting that already at this time transport properties
viscosity and Reynolds number of the hot and compressed nuclear matter
were studied
\cite{bc8788},
via energy and mass scaling of collective flow data.

In the 1990s flow measurements continued to develop. With
increasing energy the dominance of directed flow decreased,
due to the increasing longitudinal momentum component and
decreasing collective transverse component of the flow.
At the same time the azimuthal asymmetry of collective
flow became more dominant and stronger {\it elliptic flow}
was detected with increasing beam energies.  Simultaneously
new forms of collective flow parameterizations were introduced,
$ v_1, v_2, v_3, ...$ etc.  This parametrization provided some
advantages and was more systematic in describing the azimuthal
symmetries and asymmetries of the reaction. Still, it
led to some side-effects, which complicated the subsequent 
development.

First of all the longitudinal momentum distribution was not 
analysed with the same precision and systematic way.
This is in part due to the experimental conditions, where 
at increasing energies the measurement of the particle emission
at a larger rapidity range between the projectile rapidity, $y_P$,
and the target rapidity, $y_T$, became increasingly
difficult (although recently a similar harmonic expansion
in terms of Chebyshev polynomials was also introduced
\cite{BzTe13}). Due to the experimental constrains this
is unfortunately not yet used in experimental analyses.

The other side effect of this development that the symmetry axes 
of the collision, i.e. the event-by-event direction of the 
impact parameter vector ($x$-direction) was not performed in 
all experiments, neither the precise determination of the 
CM of the participant system, which can fluctuate
both in the beam and transverse directions. E.g. in ref.
\cite{AL2013d} 
the participant CM. was not determined (which could be 
done by using the method proposed in ref. 
\cite{Eyyubova})
neither in the transverse nor in the longitudinal direction 
where the fluctuation of the CM. rapidity is much larger.
This may have resulted in unexpected artifacts.

These developments and experimental difficulties led to the
dominance of the elliptic flow (as well as other even $v_n$
studies where an "Event plane" could be determined by "cumulative"
two particle correlations, without determining the 
Reaction Plane (i.e. the impact parameter vector's direction)
precisely. This also led to some confusion in analysing
flow patterns arising from initial state collision symmetries
and from different random fluctuations. We will discuss this 
problem later in more detail.

The collective flow in form of elliptic flow remained a sensitive
signal of the changes in EoS and of all matter properties.
It was predicted that the signature of QGP formation and the
related softening of the EoS, modifies the $v_1$ and $v_2$ data
and lead to an appearance of the "third flow component" or
"anti-flow",
\cite{cr99}
which was observed a few years later together with several other signals
of QGP formation. In many of these observations the collective flow
played important direct or indirect role.

\subsection{Causes of the Dominance of Fluid Dynamics}
\label{causes}

With increasing energies, especially at RHIC, the collective flow 
effects became increasingly dominant. This was to be expected as 
the hadron multiplicity grew above 5000 in these reactions, and
the matter more and more behaved like a continuum. On the other hand
questions of the fundamental stability and the role of dissipation,
especially of viscosity gained increasing attention again.

Collective laminar flow is based on momentum correlations among
numerous neighboring particles. This requires momentum exchange
among neighboring fluid elements. This leads to the reduction
of local flow velocity differences, and these differences are 
dissipated into heat. Such a process is described by the viscosity
of the fluid.  If viscosity is too small the flow becomes turbulent.
If viscosity is too large, dissipation becomes too large and
flow patterns characterized by different flow velocities at
different locations are dissipated away.

The stability of the flow requires a certain minimum level of 
viscosity. This stability against turbulent instabilities is
characterized by the dimensionless, Reynolds number:
$$
Re = l_1 u_1 / \nu
$$
where $l_1$ is the characteristic length scale of the
flow pattern, $u_1$ is the characteristic flow velocity,
$\nu = \eta/\rho$ is the kinetic viscosity, $\eta$ is the 
shear viscosity, and $\rho$ is the mass (or energy) density.

In an ideal (or perfect) fluid any small perturbation 
increases and leads to turbulent flow. For stability 
sufficiently large viscosity and/or heat conductivity 
are needed! 
$$
   Re \le  1000  -  2000
$$
Calculations are also stabilized by numerical viscosity!
(Even if the equations which are solved are formally 
describing a perfect fluid with zero viscosity.)

This stability condition is originally applied to mechanical
stability, for processes around or above the sound speed.
Highly energetic or high temperature phenomena present an
additional challenge.  Here it was realized recently the
decisive role of radiative energy and momentum transfer,
thus the application of relativistic fluid dynamics is
vital. Thus apart of heavy ion reactions these stability
considerations are vital in rocket propulsion, energetic implosions,
fission- and fusion reactions, etc. 

Numerous recent theoretical studies, especially by  
D. Molnar \cite{MolnarD08}, 
U. Heinz \cite{Heinz08},  
et al., estimated the value of viscosity in 
ultra-relativistic heavy ion reactions at RHIC energies as 
$ \eta \ \approx\  50 - 500 {\rm MeV/fm}^2{\rm c}$ thus
$ Re\ \approx\  10 - 100 $.
Based on experimental results in the  energy range of 
50 - 800 MeV/nucleon in the 1980"s using scaling analysis 
of flow parameters, the Reynolds number was obtained as
$ Re \approx 7 - 8 $.
\cite{bc8788} 
This was a result for more dilute,
and so, more viscous matter.

In both cases $\eta/s \approx 1$ (0.5 - 5). This is a value 
large enough to keep the flow laminar in Heavy Ion Collisions!

On the other hand the surprising dominance and strength of the 
collective flow phenomena indicate that the flow is near to
perfect in heavy ion reactions. About half of all available 
energy appears in form of collective flow at and below the transition
to QGP (see e.g. \cite{BCLS94} Fig. 6). 
Thus, only a small
fraction of energy could be dissipated away, especially as the
phase transition itself contributes to dissipation and entropy
increase also. Thus, this is almost a paradox, the very strong
flow also indicates that the viscosity is small, and so
the viscosity must be sufficiently small and sufficiently large
at the same time, to satisfy all requirements and observations.

Another comment is relevant here: The collective flow must develop
mainly in the quark gluon phase. There are strong indications
that hadronization and freeze out are happening simultaneously
in heavy ion reactions \cite{Csorgo94,CsMi95}. 
This is indicated by the large abundance of multi-strange baryons and 
by the small freeze out size extracted from two particle correlation
data. Finally the scaling of flow with the quark number of
observed hadrons also indicates that flow develops mostly in the 
quark-gluon phase. Thus, the physical features which considerably
influence flow phenomena at RHIC energies (like viscosity and EoS),
should be primarily the features of QGP!

Since usually the bulk viscosity is small compared to the shear 
viscosity, the dimensionless ratio of (shear) viscosity to entropy (disorder) 
$\eta/s$ is a good way to characterize the intrinsic ability of a substance to 
relax towards equilibrium.

The recent developments address both the EoS and the transport properties
of the extreme matter formed in heavy ion reactions. These aspects 
were addressed recently \cite{CsW13Koly}, and a part of the issues 
discussed here were mentioned at this conference. 

In a recent paper Kovtun, Son and Starinets have 
shown that certain  field theories, 
that are dual to black branes in higher space-time 
dimensions, have the ratio $\eta/s = 1/4\pi$ \cite{Kovtun2005}.  
Interestingly, this bound is obeyed by ${\cal N}=4$ 
supersymmetric $SU(N_c)$ Yang-Mills theory in the large $N_c$ limit 
\cite{SUSY}.  
They "speculated"  that all substances obeyed this bound, and 
argue that this is a lower limit especially for such 
strongly interacting systems where up to now there is no reliable 
estimate for viscosity, like the QGP.  According to the authors: 
the viscosity of QGP must be lower than that of classical fluids.

Recent advances in the study of the collective flow properties, have
a wide spread of directions. Due to the low viscosity of Quark-Gluon
plasma near to the phase transition threshold
\cite{Kovtun2005,CKM2006}, both significant 
fluctuations may develop
\cite{KMS12-13}
and new Global Collective instabilities
may occur, as turbulence in peripheral reactions. The precise 
analysis of these effects would require the experimental separation
of these effects as well as the theoretical study of these
effects separately, and then their possible interaction and interference 
in the observables.
The necessity of this separation was pointed out recently
\cite{INPC13}, and we will elaborate this subject in more detail
here.

\begin{figure}[ht]  
\centering
\hskip -2cm
\includegraphics[width=9cm,clip]{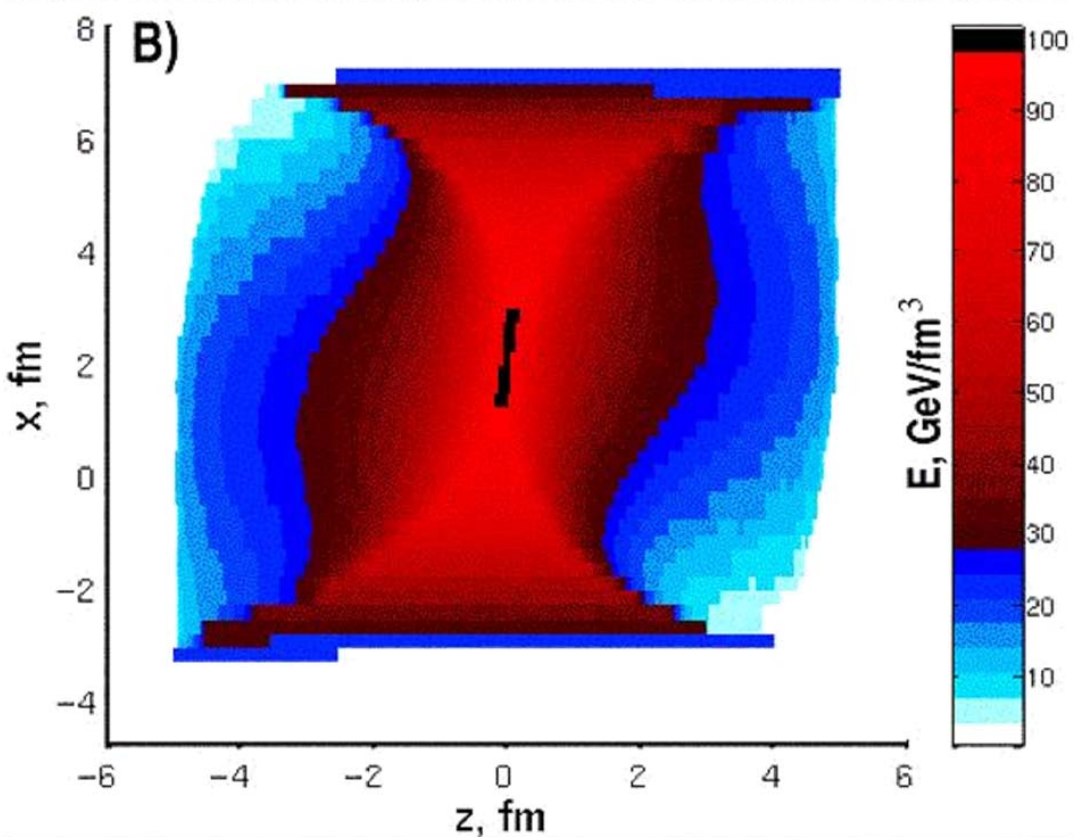}   
\caption{
The initial state energy density distribution
shown in  the Reaction Plane, in the  $[x,z]$, plane. This 
initial state is constructed based on a Glauber model, via
fire-streaks, which extend longitudinally. This
extension is slowed down by the attractive, chromo-electric, coherent
 Yang-Mils fields. The resulting string-rope tension
is smaller when we have less color charges at the end of the
streaks, and this results in a longer streaks and smaller energy 
density at the top (projectile) and bottom (target) sides.
The top and bottom layers move into the opposite directions in the
CM frame. The central 
streaks that stopped stronger, start a 1D Riemann scaling expansion.
This initial state, in contrast to many oversimplified ones, conserves 
energy, momentum, angular momentum, and shows
initial vorticity and longitudinal shear. From ref. \cite{MCs001,MCs002}.
}
\label{fig-1}     
\end{figure}
Another, recent problem is the formation and study of
realistic 3+1D initial states in fully realistic description,
without unrealistic simplifying assumptions. Here, Fluctuations
and Global Collective flow should also be separated, and the 
Global Collective initial state model should reflect all symmetries
of a heavy ion reaction. This is still not always the case.

In numerous fluid dynamical models, which use the $x, y, \eta, \tau$
coordinates, it is easy to assume uniform longitudinal Bjorken scaling
flow, so that $v_\eta = 0$, and it is done frequently even in 3+1D models,
this eliminates immediately the longitudinal shear flow, and the 
arising vorticity. When the longitudinal momentum distribution is
uniform in the transverse plane or if it is symmetric around the
collision's $z$-axis the initial angular momentum is lost, which
disables the description of many fundamental phenomena, and 
also violates the conservation of angular momentum.
For a realistic model the longitudinal ends of the initial
state (in terms of $z, t$ or $\eta, \tau$), should not
exceed the projectile and target rapidities, rather the recoil
and the deceleration caused by the other colliding nucleus 
should be also taken into account. 
Only a few models satisfy fully, all conservation laws, and
we will discuss the construction of realistic initial state
configurations for the Global Collective flow component.

There are few realizations where the conservation laws are
fully satisfied. The models generating the initial state from
a realistic molecular dynamics or cascade model may reach 
states close to equilibrium, and the smooth average of such
states can serve as an realistic 3+1D initial state.
One can also construct a good analytic initial state by
taking into account of all symmetries and all conserved quantities
and their conservation laws. Such an initial state is described
in \cite{MCs001,MCs002} and presented in Fig. \ref{fig-1}.

\section{Splitting of Global Collective Flow and Fluctuations}
\label{sec-1}
The high multiplicities at high energy heavy ion collisions have enabled us
to study fluctuations and the distribution of the azimuthal harmonic 
components. Due to traditional reasons the azimuthal distributions are 
parametrized in terms of cosine functions and a separate event-by-event
fitted Event Plane azimuth, which did not correlate with the Reaction
Plane and had nontrivial correlation among the Event Planes of the 
different harmonic components.  

The non-fluctuating Global Collective (background)
\footnote{
In ref. \cite{FW13nov} this component is called the background contribution.
}
flow, if the event-by-event
center of mass and Reaction Plane are identified, can be written in the
form
\ba
\frac{d^3N}{dydp_td\phi} = \frac{1}{2\pi}\frac{d^2N}{dydp_t} \left[ 1 
+  2v_1(y-y_{CM},p_t) \cos(\phi-\Psi_{RP}) + \right. 
\nonumber \\ \hskip 4.5cm 
\left. 2v_2(y-y_{CM},p_t) \cos(2(\phi-\Psi_{RP})) + \cdot\cdot\cdot \  \right]
\ea
where
$\Psi_{RP}$ \ \   and  \ \   $y_{CM}$   can be determined 
experimentally event-by event, as described in \cite{Eyyubova}.

The Fourier components are functions of $p_t$ and rapidity
$y$. Most data are collected for $v_2(y,p_t)$ at and near
CM rapidities, and as function
of $p_t$ . These data are quite well fitted already 
by early simple FD models, even by those,
which did not calculate the fluid's dynamics in the $z$ direction, 
rather these assumed that the longitudinal part of the
expansion is described by the Bjorken's scaling FD model.

Notice that due to fluctuations the event-by event CM fluctuates 
strongly, in the beam direction  $y_{CM}$ fluctuates
due to the large rapidity difference between the projectile and 
target, and due to the event-by event CM fluctuation, also 
the event-by-event azimuthal angle of the Reaction 
Plane, $\Psi_{RP}$, will be modified. This second effect was taken
into account in ref. \cite{AL2013d}, without referring to
 \cite{Eyyubova},
but the stronger longitudinal fluctuations were not studied, and
were considered just as "dipole like initial fluctuations".

In contrast to the above formulation, fluctuating flow patterns are
analysed by using the ansatz
\ba
\frac{d^3N}{dydp_td\phi} = \frac{1}{2\pi}\frac{d^2N}{dydp_t} \left[ 1 
+  2v_1(y,p_t) \cos(\phi-\Psi_1^{EP}) + \right. 
\nonumber \\ \hskip 4.5cm
\left. 2v_2(y,p_t) \cos(2(\phi-\Psi_2^{EP})) + \cdot\cdot\cdot \  \right] ,
\ea
which is adequate for exactly central collisions where the
Global Collective flow does not lead to azimuthal asymmetries. Here
$ \Psi_n^{EP}$  maximizes $v_n(y,p_t)$ in a rapidity range, and both 
$\phi$ and $ \Psi_n^{EP}$ are measured in the laboratory (collider) frame.

If this formulation is used for peripheral collisions the
analysis is rather problematic, because patterns arising from
Global Collective flow
and from fluctuations are getting mixed up. This is actually
also true in central collisions with spherical or cylindrical 
symmetry but there
the separation is more subtle, and it does not show up directly
in the azimuthal flow harmonics. Still in special model calculations
fluctuations in the transverse plane were studied, and  
Global Collective flow (background flow) was separated from
fluctuations \cite{FW13nov}.

Here we show that the ansatz of flow analysis can be reformulated in a way,
which makes the splitting or separation of the Global Collective
flow from Fluctuations easier. This can be done 
based on the final components of given types of symmetries
arising from the symmetries of the initial states of peripheral 
heavy ion collisions.
This formulation is also an ortho-normal series expansion 
for both $\phi$-even and $\phi$-odd   functions. 

Considering the relation
$\cos (\alpha{-}\beta) = 
\cos\! \alpha \cos\! \beta + \sin\! \alpha \sin\! \beta$,\  
we can write each of the terms of the harmonic expansion into the form
\[
v_n \cos[n(\phi - \Psi_n^{EP})] = 
v_n \cos(n \Psi_n^{EP})  \ \cos(n \phi)  +
v_n \sin(n \Psi_n^{EP})  \ \sin(n \phi) .
\]
If we consider that the unique reaction plane, 
$\Psi_{RP}$ can be determined event-by-event experimentally 
also  \cite{Eyyubova}, we can introduce 
$\Phi_n^{EP} \equiv  \Psi_n^{EP} - \Psi_{RP}$ and 
$\phi' \equiv \phi - \Psi_{RP}$,
so that from these data we get 
$\Psi_n^{EP} = \Phi_n^{EP} + \Psi_{RP}$.
Here $\phi'$ is the azimuth angle with respect to the Reaction Plane.
Now we can also define the new flow harmonic coefficients by 
$$
^cv'_n \equiv v_n \cos(n (\Psi_n^{EP} )) \ \ {\rm and} \ \ 
^sv'_n \equiv v_n \sin(n (\Psi_n^{EP} )) \ ,
$$
and we get for the terms of the harmonic expansion
\be
v_n \cos[n(\phi  - \Psi_n^{EP})]=\
v_n \cos[n(\phi' - \Phi_n^{EP})]\ 
=\ ^c\!v'_n \cos(n \phi') +\ ^s\!v'_n \sin(n \phi') . \ \ 
\label{eq-1}
\ee
Thus we have reformulated the azimuthal angle harmonic expansion,
which was given originally
in terms of cosines and Event Plane angles for each
harmonic component, to both sines and cosines in the Reaction Plane 
as reference plane and the corresponding new coefficients
$ ^c\!v'_n =  ^c\!v'_n (y-y_{CM},p_t)$ and
$ ^s\!v'_n =  ^s\!v'_n (y-y_{CM},p_t)$.
These can be obtained from the measured data, $v_n$, $\Psi_n^{EP}$,
$\Psi_{RP}$ and $y_{CM}$ directly.

As the Global Collective flow in the configuration space 
has to be $\pm y$ symmetric, all the coefficients of 
the $\sin(n \phi')$ terms should vanish:
$ ^s\!v'_n = 0$.
These symmetry properties provide a possibility to separate the
fluctuating and the global flow (background flow) components.
\footnote{
In ref. \cite{FW13nov} for the longitudinal motion uniformly the
Bjorken scaling flow approximation was assumed, which is inadequate
to describe the odd $(y-y_{CM})$ components. Thus this analysis is
limited in the possibility of separating the two components.
This is already included in the ansatz of the assumed 
distribution function, $\delta f_i$ in eq. (2.9) where longitudinal
fluctuations were excluded and only transverse fluctuations were
studied.}

This form has the advantage that in peripheral collisions the
Global Collective (not fluctuating) flow component, 
$^c\!v'_n$ for odd harmonics have to be odd functions of $(y-y_{CM})$, 
while 
for even harmonic components have to be even function of rapidity,
 $(y-y_{CM})$.
Let us now introduce the rapidity variable ${\bf y} \equiv y-y_{CM}$.

When the new coefficients
$ ^c\!v'_n =\  ^c\!v'_n ({\bf y},p_t)$ and
$ ^s\!v'_n =\  ^s\!v'_n ({\bf y},p_t)$, are constructed,
we can conclude that $ ^s\!v'_n$ can be due to fluctuations
only. Furthermore for the Global Collective flow, 
$^c\!v'_n ({\bf y},p_t)$ must be an even (odd) function of
${\bf y}$ for even (odd) harmonic coefficients.
Due to the fluctuations this is usually not satisfied and one has to
construct the even (odd) combinations from the measured data.
These represent then the Global Collective component, while the
odd (even) combination will represent the Fluctuating
component. We obtain the Fluctuating component by subtracting the
Global Collective component from the Total (mixed) flow harmonic term:
\ba
v_{n\frac{even}{odd}}^{Coll.} \cos[n(\phi  - \Psi_n^{EP})] &=&
\frac{1}{2}\left[^c\!v'_n ({\bf y},p_t) \pm\ ^c\!v'_n (-{\bf y},p_t) \right]
\cos(n \phi') \\
v_{n\frac{even}{odd}}^{Fluct.} \cos[n(\phi  - \Psi_n^{EP})] &=&
\frac{1}{2}\left[^c\!v'_n ({\bf y},p_t) \mp\ ^c\!v'_n (-{\bf y},p_t) \right]
\cos(n \phi') 
\nonumber \\
 & & \phantom{\frac{1}{2}\left[\right.} + ^s\!v'_n ({\bf y},p_t)\ \sin(n \phi')
\ea

This separation provides an upper limit for the magnitude
of the Global Flow component, because the fluctuations 
may in some events show the same symmetries as the Global
Collective flow. On the other hand, for the Fluctuating
component, $^s\!v'_n$, provides an upper limit, because this component
cannot be caused by the Global Collective flow.
A last essential guidance may be given by the conditions
that the fluctuations must have the same magnitude for sine
and cosine components as well as for odd and even rapidity
components.

The evaluation of experimental results this way may provide 
a better insight into both types of flow patterns. Furthermore,
these can also help judgements on theoretical model results, and the 
theoretical assumptions regarding the initial states.

Other experimental methods, like two particle correlations
\cite{CVW13} or polarization measurements \cite{BCW2013},
may take advantage of this splitting of flow pattern 
components also.

\subsection{Negative $v_1(p_t)$}
\label{sec-2.1}

The Global Collective directed flow, $v_1$, becomes negative
due to the softening of the EoS in Quark-gluon Plasma
\cite{cr99,antiflow}. This shows up at CERN SPS,
RHIC and LHC energies, and was first predicted in fluid dynamical
calculations in 1994, see \cite{BCLS94} Fig. 8, and in 1995
\cite{RPMetal19}. The softening effected $v_2$ also as predicted e.g. 
in RQMD calculations \cite{Sorge99}.
This softening makes $v_1({\bf y})$ negative at small positive
rapidities, and thus at the these rapidities $v_1({\bf y}, p_t)$
will also be negative. The rapidity integrated  $v_1(p_t)$ at the 
same time should vanish as the Collective directed flow, $v_1$, 
should be an odd function of rapidity. Furthermore in any case 
due to transverse momentum conservation  $\langle v_1(p_t) \rangle = 0$
\cite{ZFBF01}. In Global Collective flow models, e.g. in
fluid dynamical models without random fluctuations, 
the integrated $v_1(p_t)$ should vanish, while the 
symmetrized $v^S_1(p_t)$ \cite{hydro1} can be finite and usually
still positive. 

On the other hand, recent measurements yield negative  $v^S_1(p_t)$  
values at low rapidities, $p_t < 1.2 - 1.5$Gev/c
\cite{Jia-ATLAS-12,ALICE-PLB-12,AL2013d}. 
The same is observed
in model calculations both in fluid dynamics \cite{GaleEtal-12} 
and in molecular dynamics \cite{ZFBF01}
with random fluctuating initial conditions. This is not unexpected.

Still there is a problem. When the Global Collective flow and 
Fluctuating flow are not separated and the CM is not identified
the contributions of Events with different unidentified
CM points can lead to negative $v^S_1(p_t)$ also, as it can be 
clearly seen from eqs. (2) and (3) of ref. \cite{hydro1}. This
is a consequence of not identifying the EbE CM of the 
participant system, which is also fluctuating, but this
fluctuation has different physical grounds then e.g. critical
fluctuations arising from the phase transition.

The main cause of the longitudinal CM fluctuations is
that the target and projectile spectators must not be
identical and so, their momenta are also different. At the same
time this fluctuation does not influence
the nonlinear dynamical evolution of the participant system, 
which arises from the random initial configuration, the 
critical dynamics, and the randomness in the freeze out
process.

The interference of the global collective and the fluctuating
dynamics is well represented by the detailed results in ref.
\cite{Jia-ATLAS-12}.

In order to extract the physical features of the QGP based
on the observed fluctuations it is of utmost importance to separate
the Global Collective flow components from the Fluctuating flow
components.

Gyulassy et al. \cite{GLVB14} claim that
recent low $p_t<2$ GeV azimuthal correlation data from the beam energy 
scan (BES) and D+Au at RHIC/BNL and the especially the surprising low 
$p_t$ azimuthal $v_n(p_t)$ in p+Pb at LHC challenge
long held assumptions about the necessity of perfect fluidity 
(minimal viscosity to entropy, $~ 1/4\pi$) to account for azimuthal 
asymmetric "flow" patterns in A+A.
The work discusses basic pQCD interference phenomena from beam jet 
color antenna arrays that may help unravel these puzzles without 
requiring perfect fluid hydrodynamic or CGC Glasma diagrams, 
but only LO Feynmann diagrams.

First we have to comment that comparing p+p and peripheral Pb+Pb 
data is not trivial, as most of the time equal multiplicity collision
are compared, where the geometry of the two types of collisions is
very different. 

Furthermore the Global Flow and the Fluctuating Flow dynamics
are not analyzed separately, thus it is unclear what these
comparisons refer to.

Finally the $v_1$ measurements, which are separately mentioned
in ref. \cite{GLVB14}, their methods, and the results
are problematic and therefore avoided in most experimental and 
theoretical publications. As a matter of fact similar problems 
appear in case of all odd flow harmonics also, but here the
fluctuating component is dominant so the effect of separating
the background Global Collective flow component does not
cause a large difference.

$v_1(y)$ observations show a central antiflow slope,
$\partial v_1(y) / \partial y$, which is gradually decreasing 
with increasing beam energy \cite{AL2013d}: 
$$
\frac{\partial v_1(y)_{odd}}{\partial y}
=\left\{ \begin{array}{llrl}
- 1.25\%\ &  {\rm for}\ & 62.4\ {\rm GeV} &{\rm (STAR)} \\
- 0.41\%\ &  {\rm for}\ & 200.0\ {\rm GeV} &{\rm (STAR)} \\
- 0.15\%\ &  {\rm for}\ & 2760.0\ {\rm GeV} &{\rm (ALICE)} \\
 \end{array} \right.
$$
This decrease is partly due to the smaller increase of the
pressure and the caused transverse motion. In addition the
forward rotation of the antiflow peak with increasing
angular momentum \cite{hydro1} leads to
a decrease also as the peak approaches the turnover
point where the antiflow peak would turn over to directed flow.
The non-identification of the participant system's CM leads to a 
strong smoothing and decrease of the  $v_1(y)$ peak \cite{hydro1},
but the experimental result \cite{AL2013d} indicates that 
the peak is still in the antiflow direction, and the predicted turn
over to directed flow did not happen. This shows that 
the rotation at the present 2.76 TeV energy is not sufficient
to reach the turnover point before freeze-out. However, at the 
double beam energy with double angular momentum the turnover
may take place.

Returning the alternative suggestion of Gyulassy et al. \cite{GLVB14} 
of creating
azimuthal asymmetry in microscopic processes in LO Feynmann diagrams,
these may have a role in p+p or eventually in p+A collisions,
however, as discussed above the azimuthal and longitudinal 
dynamics is strongly connected in A+A collisions with substantial
Global Collective flow processes. In Pb+Pb collisions there is 
to long a distance between the sources/minijets, which according 
to \cite{GLVB14} shall produce the $v_2$ (and higher even) flow harmonics 
from their Gluon correlation/interference. The modifications are 
independent of pseudorapidity, thus will have minor or no contributions to odd
flow harmonics.  In Pb+Pb it is hardly feasible 
that these processes will be correlated with the Reaction Plane and the 
participant CM, thus only the Fluctuating component will be
influenced by these azimuthal correlations of microscopic origin.

Reference \cite{GLVB14} points out the negative $v_1(p_t)$ 
at low transverse momenta.
We discussed this effect in this subsection in general by comparing the 
Global Collective and Fluctuation contributions, and the effect of 
non-identification of the participant CM. The processes discussed
in \cite{GLVB14} can thus contribute to the Fluctuating component of the 
flow. This, however has many other origins: Initial configuration fluctuations, 
Langevin fluctuations in critical dynamics, fluctuations arising
from the hadronization and freeze out, etc. The separate analysis of these
numerous processes is particularly difficult experimentally. The reasonable
first step of such an analysis is to extract the non-Fluctuating global
collective flow, and in the next step one 
may distinguish the different origins
of the fluctuating flow. 

We conclude that ref. \cite{GLVB14} will not contribute to the study of 
the negative $v_1(p_t)$ at low transverse momenta, other than the
suggested processes contribute to the fluctuation flow components 
just as the other causes of fluctuation. With increasing system size the
role and observability of these microscopic processes must decrease.

\section{The Initial State}
\label{sec-2}
As we have shown in the introduction the 
Initial State can be constructed in a way
such that all conservations laws are satisfied, and 
no simplifying assumptions are used, which would violate the
conservation laws. In addition there are other principles 
like causality which should also be satisfied by the initial 
state. 

A frequent simplification in  $x, y, \eta, \tau$
coordinates, is to assume uniform longitudinal 
Bjorken scaling flow (this leads to a simple separable initial state
distribution function), and in order to satisfy the angular 
momentum conservation at different transverse points the
energy density or mass distribution is made such that on the
projectile side a substantial part of the mass is at rapidities
exceeding the target and projectile rapidity 
\cite{schenke,Bozek,Adil}. 
In most cases this leads to acausal distributions
where part of the matter is situated beyond the target and projectile
rapidities. This acausality is corrected by Karpenko et al.
\cite{Karpenko} by cutting the distributions at the target and
projectile rapidities. Still the attractive chromo field is not taken into
account in this approach, which would limit the initial
limiting rapidities by up to 2.5 units of smaller rapidities on each side
\cite{CK1984,CK1985,Mishu02}.

Furthermore, the Bjorken scaling flow approach eliminates
any possibility for initial shear flow and vorticity, which is a 
dominant source of simple flow patterns and of strong and visible 
instabilities in classical physics, like rotation and turbulent
Kelvin Helmholtz Instability. Apart of the semi-analytic initial
state model mentioned in the introduction other initial state 
models exist, which satisfy all conditions of a realistic
initial state. First of all initial state molecular dynamics and
multi-particle cascade models which satisfy all conservation laws,
boundary conditions and causality, will provide realistic 
Global Collective initial state as the average of many such
realistic events. Also, analytic models can be constructed based
on these principles, which are different from the one mentioned in the
introduction.  

The initial uniform Bjorken scaling flow is maintained during the 
fluid dynamical development, so that the lack of shear-flow persists
in these solutions. It follows that no viscous dissipation takes
place in the longitudinal direction, which makes the model configurations
anisotropic and not very reliable in this model configurations. 
\footnote{
From the numerical point of view this initial state and this
reference frame lead to additional dynamical problems: The
longitudinal cell size is changing during the solution, while 
the transverse cell sizes remain the constant. The coarse graining 
arising from the cell sizes becomes anisotropic in the
 $x, y, \eta, \tau$ frame. As the dissipation and numerical viscosity
are proportional with the cell sizes, these will lead to increasing
anisotropy in the dissipation and in the numerical viscosity.
This leads to unwanted numerical artifacts.}

Initial State (IS) models are frequently based on molecular dynamics 
type or string models like, uRQMD, QGSM or PACIAE, where the interacting
gluon fields play a secondary role. Since 2001, recent IS models take into 
consideration strong gluon fields arising from the Color-glass Condensate
(CGC) \cite{CGC2001}, which lead to more compact and longitudinally
less extended initial spatial configurations. A few models,
\cite{MCs001,MCs002,Mishu02} 
have taken such a collective, attractive classical Yang-Mills (CYM) 
flux tube picture into account even before
\cite{GyCs86}.  
More recent models, e.g. IP-Sat and IP-Glasma 
\cite{IP-Sat03,IP-Glasma} 
use the CGC Glasma flux tube picture.
The resulting more compact initial state leads to stronger
equilibration and enhances collective dynamics, including rotation.
While the earlier models 
\cite{MCs001,MCs002,Mishu02} 
did not take into account random fluctuations the more recent
ones mostly do, using the MC-Glauber or similar schemes.
This Scheme if adequately used leads to large conserved angular momentum 
and shear flow 
\cite{Vov13}, 
although in some applications the shear flow
and the angular momentum are neglected and then
the collective flow in the odd harmonics is eliminated.

\begin{figure*}[h]
\hfill
\includegraphics[width=6.5cm]{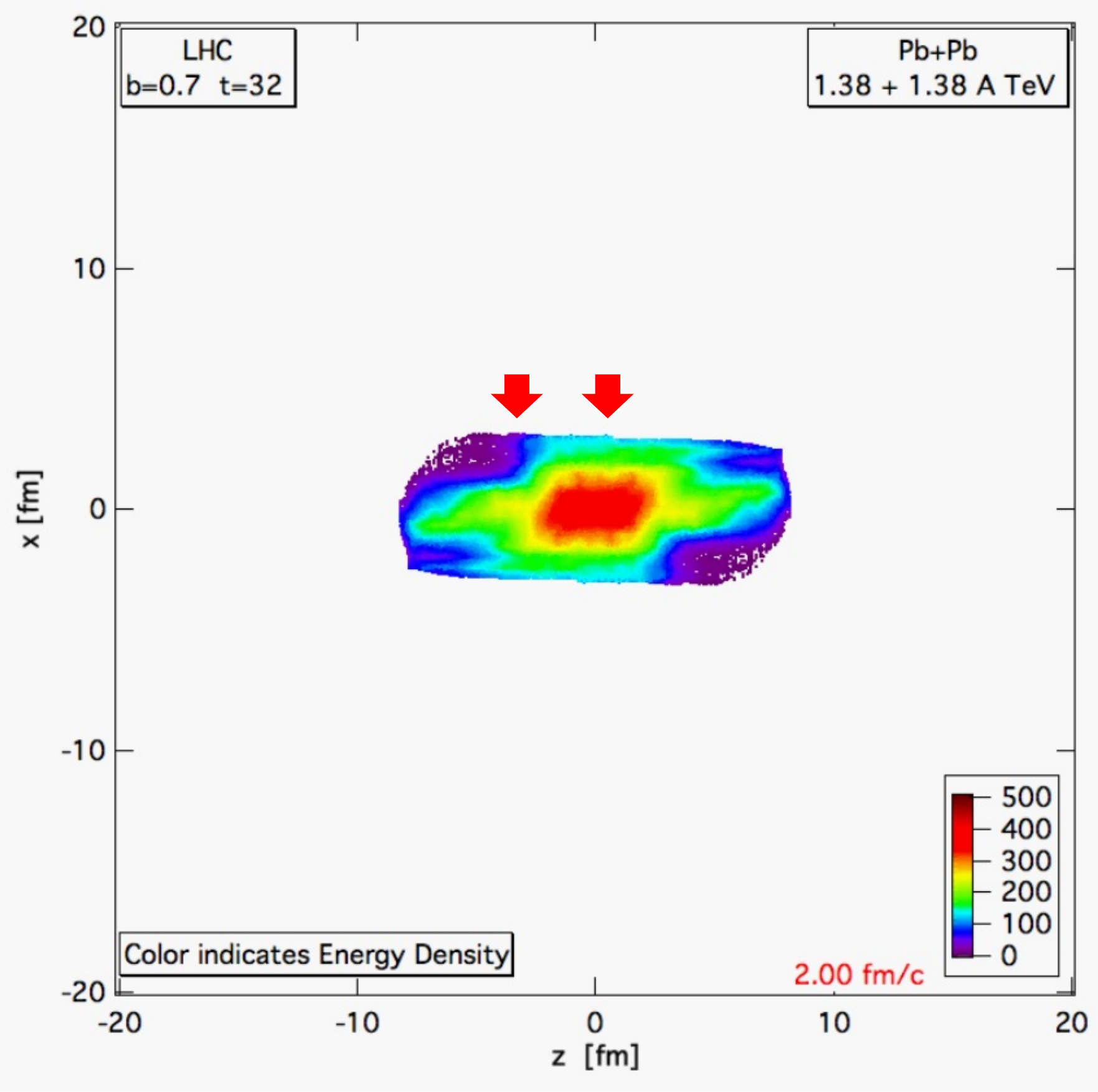}
\includegraphics[width=6.5cm]{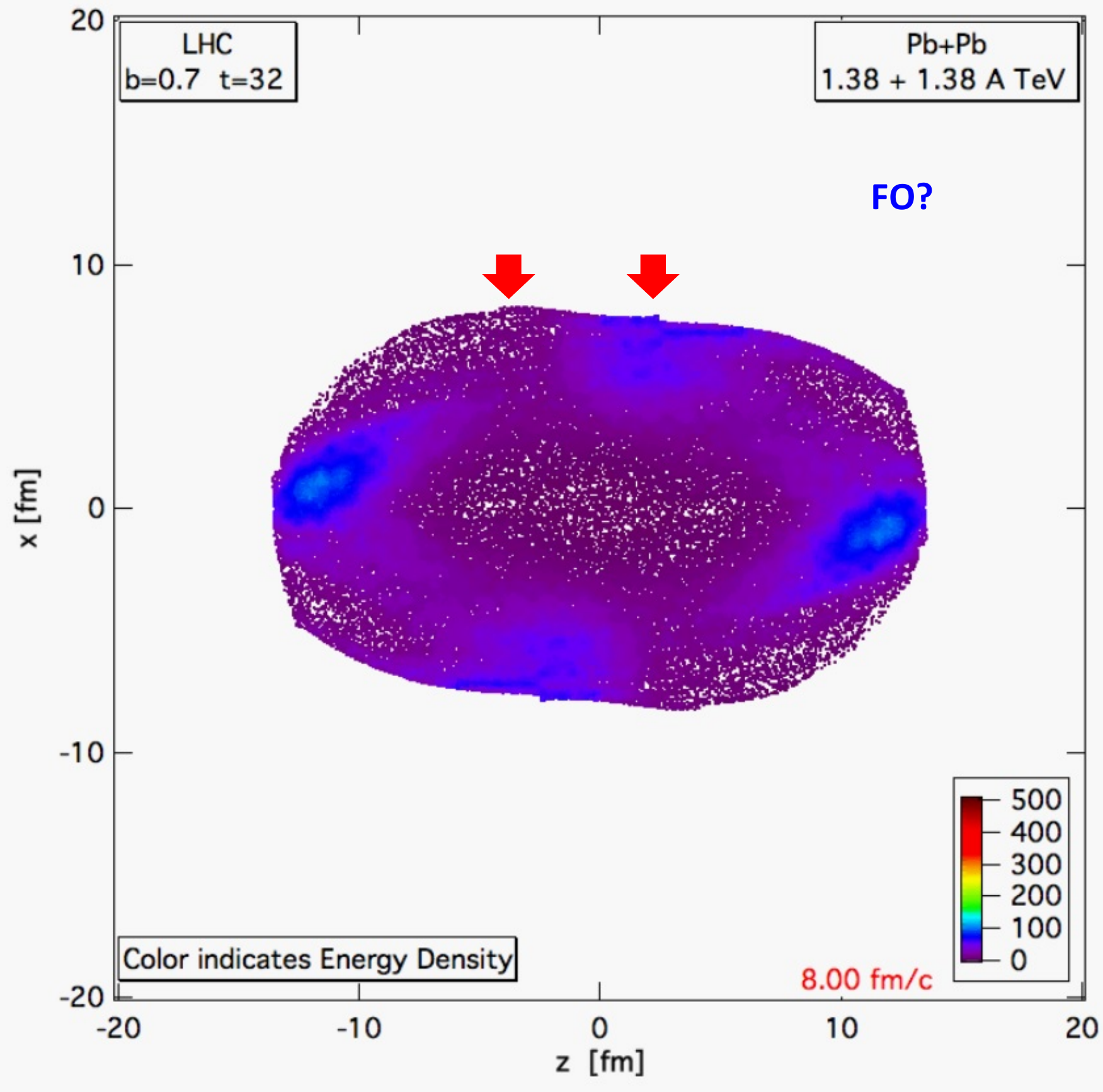}
\caption{
The rotation during the fluid dynamical evolution is indicated by the
red arrows pointing to the initial central and corner points on the
surface. The motion of these points shows the rotation of the system.
The fluid dynamical initial state is preceded
by an pre-equilibrium Yang-Mills longitudinal field theoretical model,
which took 6.25 fm/c. Thus after 2.00 fm/c fluid dynamical evolution
the length of the matter is 8.25 fm (l.h.s). The
configuration on the r.h.s. is at 8 fm/c fluid dynamical evolution,
which is 14.25 fm/c after the initial touch of the two nuclei. This is
just after the estimated freeze out time of 10-12 fm/c.
Based on ref. \cite{hydro1} and \cite{CsW13Koly}.
}
\label{fig-2}       
\end{figure*}
%

\section{New Global Collective Flow Patterns}
\label{sec-3}

As mentioned in the introduction, in collisions of finite impact parameter
at high energies we have a large angular momentum which can be as high as
J = $10^6$ $\hbar$ at LHC. 
The angular momentum is conserved,
but due to the explosive expansion of the system the angular
velocity of the participant system is rapidly decreasing, thus the
local rotation, the vorticity, decreases with time.
It depends on the balance between the expansion and the 
angular momentum if the rotation will manifest itself in observable
quantities at the Freeze out.

Due to the widespread use of the uniform longitudinal Bjorken scaling
flow in the initial condition, the rotation did not occur in
fluid dynamical model calculations, and it was studied only recently.
First it was noticed in ref. \cite{hydro1}, see Fig. \ref{fig-2}.

\begin{figure*}[h]
\hfill
\includegraphics[width=6.5cm]{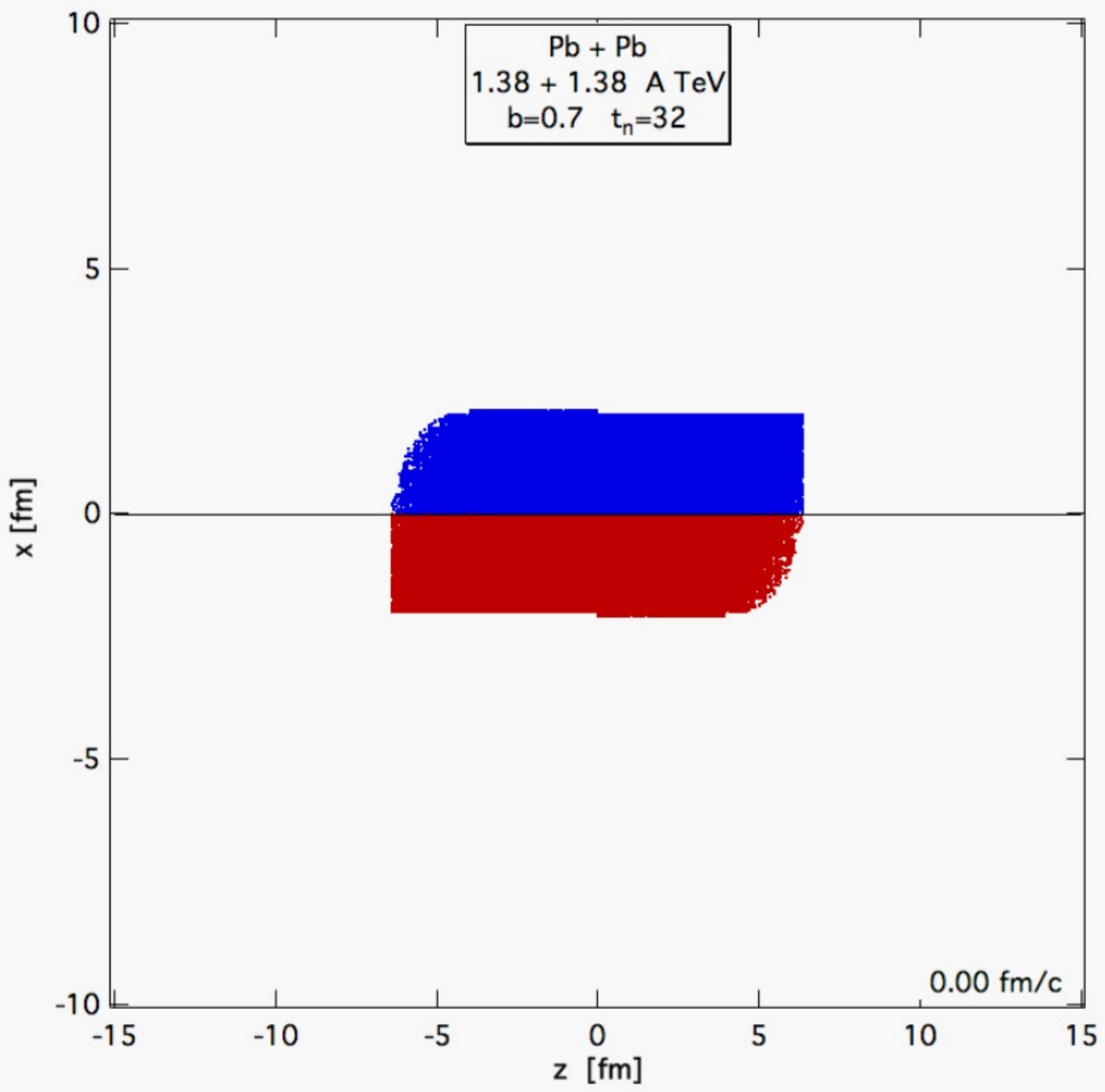}
\includegraphics[width=6.5cm]{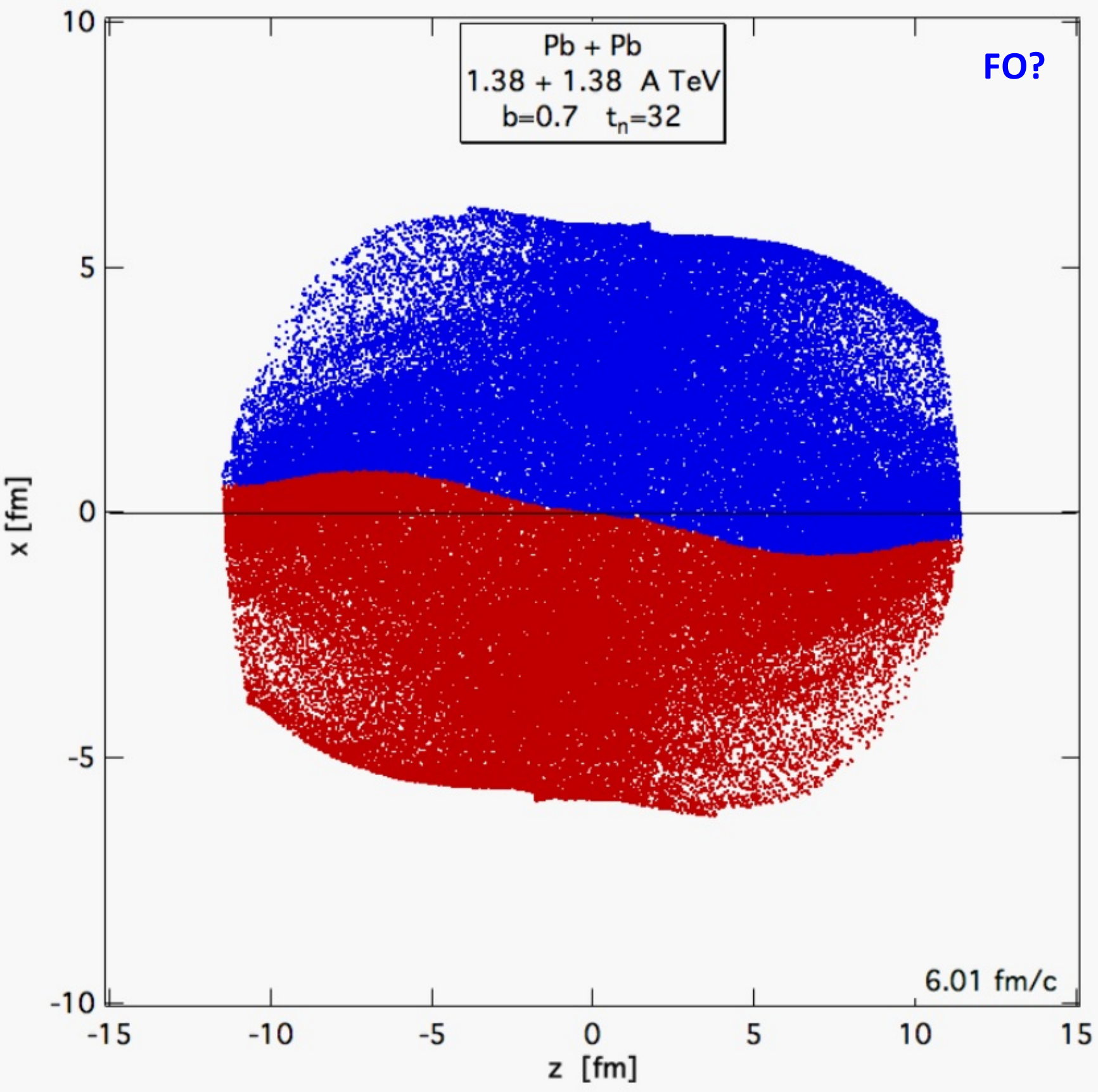}
\caption{
The fluid dynamical development of the shape of the dividing surface
between the initial top and bottom halves of the dense matter.
The developing non-linear, turbulent wave is the initial stage of
a Kelvin Helmholtz Instability.
The fluid dynamical initial state indicated by 0.00 fm/c is preceded
by an pre-equilibrium Yang-Mills longitudinal field theoretical model,
which took 6.25 fm/c indicated by the length of the dense matter. The
configuration on the r.h.s. is at 6 fm/c fluid dynamical evolution,
which is 12.25 fm/c after the initial touch of the two nuclei. This is
just around the estimated freeze out time of 10-12 fm/c.
Based on ref. \cite{hydro2} and \cite{CsW13Koly}.
}
\label{fig-4}       
\end{figure*}

This rotation acts against the 3rd flow component or antiflow, and
may decrease the measured directed flow, or even reverses the direction
from antiflow to directed flow. According the calculations  \cite{hydro1}
the $v_1$ was expected to peak at positive rapidities, but this
prediction is strongly dependent on the competition between the 
rotation and the expansion. The small amplitude of $v_1$ is difficult
to identify in the strongly fluctuating background, without 
identifying the event-by-event center of mass and Reaction Plane.

The fluid dynamical calculations with the same method showed for the
first time the possibility of the turbulent Kelvin Helmholtz Instability
\cite{hydro2}. See Fig. \ref{fig-4}.
 Stability estimates confirmed the possibility of the
occurrence if this instability, which could also be obtained in a 
simple analytic model \cite{WNC13}.

Interestingly in a recent work the holographic method is used to study
features similar to the KHI, which appear in gauge-gravity duality.
Here the dual rotating black holes are more sensitive to the asymptotic 
geometry than shearing black holes, because topologically spherical AdS 
black holes differ from their planar counterparts \cite{McInnes}.

\section{Detecting the New Flow Patterns via Polarization}
\label{sec-4}

The rotation and the turbulence have a small effect on the 
directed flow, which is weak at RHIC and LHC energies anyway, 
so alternative ways of detection should be considered.

The angular momentum in case of distributed shear flow,
shows up in local vorticity. The simplest classical
expression of vorticity 
in the reaction plane, [x-z], is defined as:
\be
\omega_y \ \equiv \ \omega_{xz} \ \equiv \ - \omega_{zx}
\equiv \ \frac{1}{2}(\partial_z v_x-\partial_x v_z)
\label{clvort}
\ee
where the $x, \ y, \ z$ components of the 3-velocity $\vec{v}$
are denoted by $v_x, \ v_y, \ v_z$ respectively. 
In 3-dimensional space the vorticity can be defined as
\be
\vec{\omega} \equiv \frac{1}{2} {\bf rot}\ \vec{v}
=  \frac{1}{2} \, \nabla \times \vec{v} \,
\label{wc}
\ee
For the relativistic case,
the vorticity tensor,
$\omega^\mu_\nu$ is defined as
\be
\omega^\mu_\nu \equiv \frac{1}{2} (\nabla_\nu u^\mu - \nabla^\mu u_\nu) ,
\ee
where for any four vector $q^\mu$ the quantity
$\nabla_\alpha q^\mu \equiv \Delta^\beta_\alpha \, \partial_\beta q^\mu =
\Delta^\beta_\alpha \, q^\mu_{,\beta}$  and
$\Delta^{\mu\nu} \equiv g^{\mu\nu} - u^\mu u^\nu$.
The relativistic 
generalization of vorticity leads to an increase of the
magnitude of vorticity \cite{CMW12}.

The local vorticity is decreasing with the expansion, but 
it is still significant at Freeze out in peripheral collisions
due to the huge initial angular momentum.
the local vorticity reaches 3 c/fm in the reaction plane
\cite{CMW12}, which is more than an order of magnitude
larger than the vorticity in the transverse plane arising
from random fluctuations \cite{Stefan}. Other similar 
collective chiral vortaic effects were predicted for lower 
energies in ref. \cite{Baznat}.

\begin{figure}[h]   
\centering
\hskip -2cm
\includegraphics[width=9cm,clip]{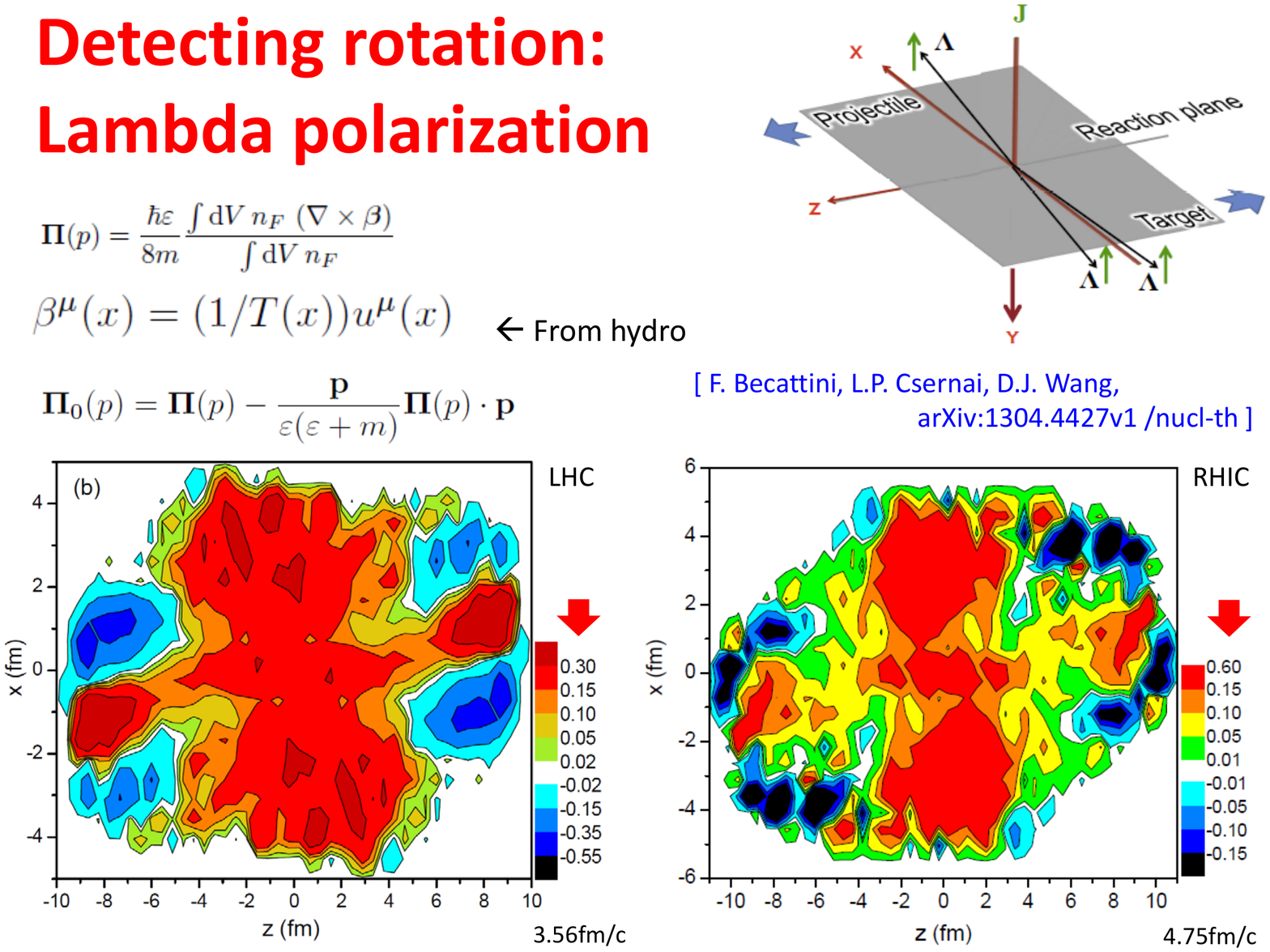}
\caption{
The $[x,z]$, Reaction Plane where the direction of the Projectile and Target
matter is indicated. The arising angular momentum, $J$, points into the
$-y$ direction. When the event-by-event center of mass  and the reaction
plane is identified this angular momentum is divided between orbital
rotation and polarization or spin. The polarization is transverse to the
motion of the $\Lambda$ and $\bar{\Lambda}$ particles and has the
same direction as the angular momentum, $J$. Thus, this polarization
may be detected in $\Lambda$ and $\bar{\Lambda}$ particles, which are emitted
into the $\pm x$ directions. From \cite{BCW2013}.
}
\label{fig-6}       
\end{figure}

This vorticity may lead to two other measurable consequences.
According to the equipartition principle for different
degrees of freedom carrying the same amount of energy
the same applies for angular momentum. Here the local orbital
rotation and the spin of the particles may equilibrate with each 
other. If equilibrium is reached by freeze out the final
polarization should have the same direction and magnitude
as the local vorticity. Interestingly, high temperature 
acts against polarization so the polarization is governed by the 
so called thermal vorticity, where instead of the four velocity,
$u^\mu$, the inverse temperature four-vector,
$$
\beta\,^\mu(x) = \frac{u\,^\mu(x)}{T(x)}\ ,
$$
is used to determine the thermal vorticity \cite{BCW2013}.
If $\beta\,^\mu$ is measured in units of $\hbar$ the thermal vorticity
becomes dimensionless.
For the polarization studies it is of utmost importance to identify 
the proper global directions in a collision event-by event. 
See Fig. \ref{fig-6}.

Without identifying the center of mass rapidity, the Reaction Plane, 
and the Projectile and target side of the reaction plane the detection of
the angular momentum and polarization is not possible and earlier
measurement at RHIC, where all azimuth angles were averaged over, gave
results where the measured polarization was consistent with zero.

\begin{figure*}[h]
\hfill
\includegraphics[width=13cm]{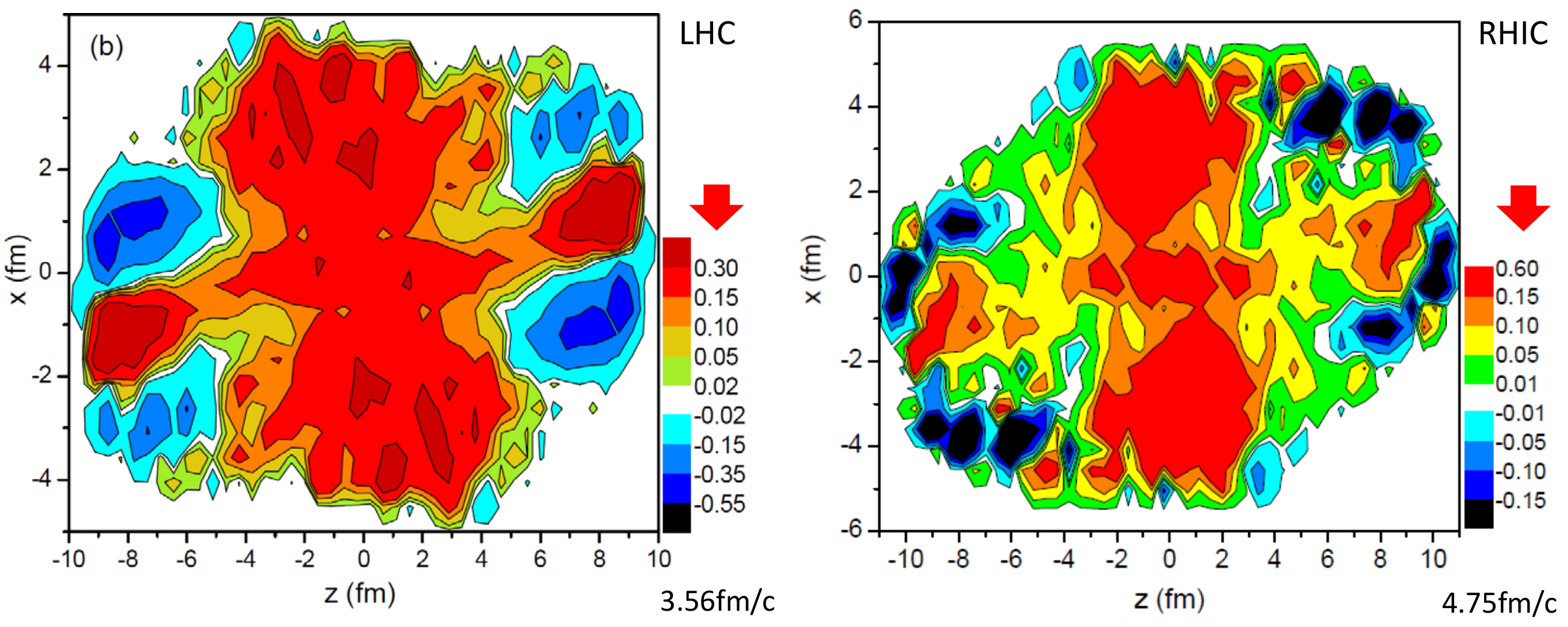}
\caption{
The thermal vorticity of the matter arising from a fluid dynamical
calculation for two different beam energies. The thermal vorticity
is inversely proportional with the temperature, which is increasing
faster than the local vorticity with increasing beam energy. Thus the
thermal vorticity at RHIC is larger. Also the side regions are cooler
and this also increases the thermal vorticity, which enhances the
polarization in due to equipartition.
Based on ref. \cite{BCW2013} and \cite{CsW13Koly}.
}
\label{fig-7}       
\end{figure*}

The $\Lambda$ particle is well suited for measuring its polarization
because its dominant decay mode is $\Lambda \longrightarrow p \, \pi^-$
and the proton is emitted in the direction of polarization. Notice 
that due to the thermal and fluid mechanical equilibration process
the polarization of  $\Lambda$s and $\bar{\Lambda}$s are the same.
This distinguishes the process from electro-magnetic polarization
mechanisms.

\begin{figure*}[h]
\hfill
\includegraphics[width=13.5cm]{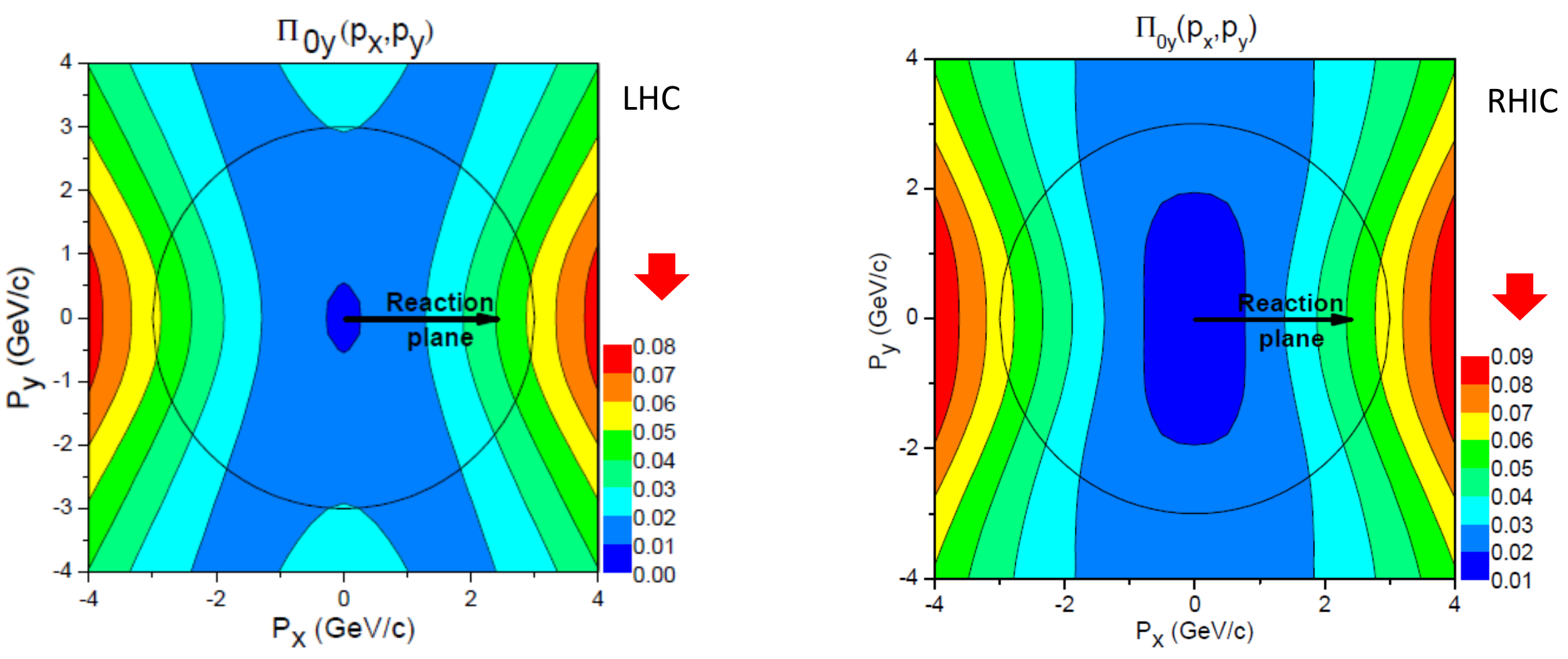}
\caption{
The polarization of the $\Lambda$ and $\bar{\Lambda}$ particles
observed in different directions and at different transverse momenta
in the transverse, $[x, y]$ plane at the event-by-event CM rapidity
for RHIC and LHC energies.
Significant polarization can be measured for particles at
larger momenta, $p_t \ge 3$ GeV/c, in the $\pm p_x$ directions.
Based on ref. \cite{BCW2013} and \cite{CsW13Koly}.
}
\label{fig-8}       
\end{figure*}

The thermal vorticity projected to the Reaction Plane is shown in 
Fig. \ref{fig-7}. The thermal vorticity is more pronounced than the
standard vorticity, at the external edges of the matter.
 where the temperature is lower. The thermal vorticity is
somewhat larger at RHIC, where the amount of data and the
available detector acceptance are larger.

The resulting polarization is shown in Fig. \ref{fig-8}. Thus for this 
measurement the determination of the proper directions of the collision
axes is vital. The polarization should be measured for $\Lambda$s emitted 
into the $\pm x$ directions, which will then be polarized in the $-y$ 
direction.

This thermal and fluid mechanical polarization would not exist if
the source, the participant system in heavy ion reactions would not
have a significant vorticity. This is realized in peripheral
heavy ion reactions, which have high initial angular momentum.
Unfortunately, even some 3+1D fluid dynamical calculations assume
oversimplified initial states where initial shear and vorticity
vanishes and these are not able to show these effects.

\section{Detecting the New Flow Patterns via Two Particle Correlations}
\label{sec-5}

The detection described in the previous section \ref{sec-4}, was sensitive
to the local vorticity. Two particle correlation measurements are
sensitive to the integrated emission from the freeze out space-time
zone, the so called "homogeneity" region, where the dominant
emission is directed toward the detection, i.e. in the (out)-direction.

Recently we proposed the Differential Hanbury Brown and Twiss method
to study the rotation of the source via two particle correlations
\cite{CVW13,CV13}.
The method is based on a simple observation, if we have a
spherically symmetric or cylindrically symmetric source with
an rotation axis, or any source which is left/right symmetric with
respect to a given "out-direction", of momentum $k$, then we 
can construct from the
usual two particle correlation function with momenta
$p_1 = k + q/2$ and $p_2 = k - q/2$:
\be
C(k, q) = \frac{P_2( k + q/2, k - q/2)}{P_1( k + q/2) P_1(k - q/2)} .
\ee
This correlation function does not depend on the direction of $k$
for static, spherically or cylindrically symmetric sources, and 
gives the same value for two, $k_+$ and $k_-$ momentum vectors
which are tilted to left/right with the same tilt angle in case 
of a left/right symmetric source with respect to $k$.
Even if the source is not static but has local motion with local 
velocities, the correlation functions have the same value if the
velocity of motion is in the radial, i.e. points in the
local out-direction.

\begin{figure}[h]  
\centering
\hskip -2cm
\includegraphics[width=9cm,clip]{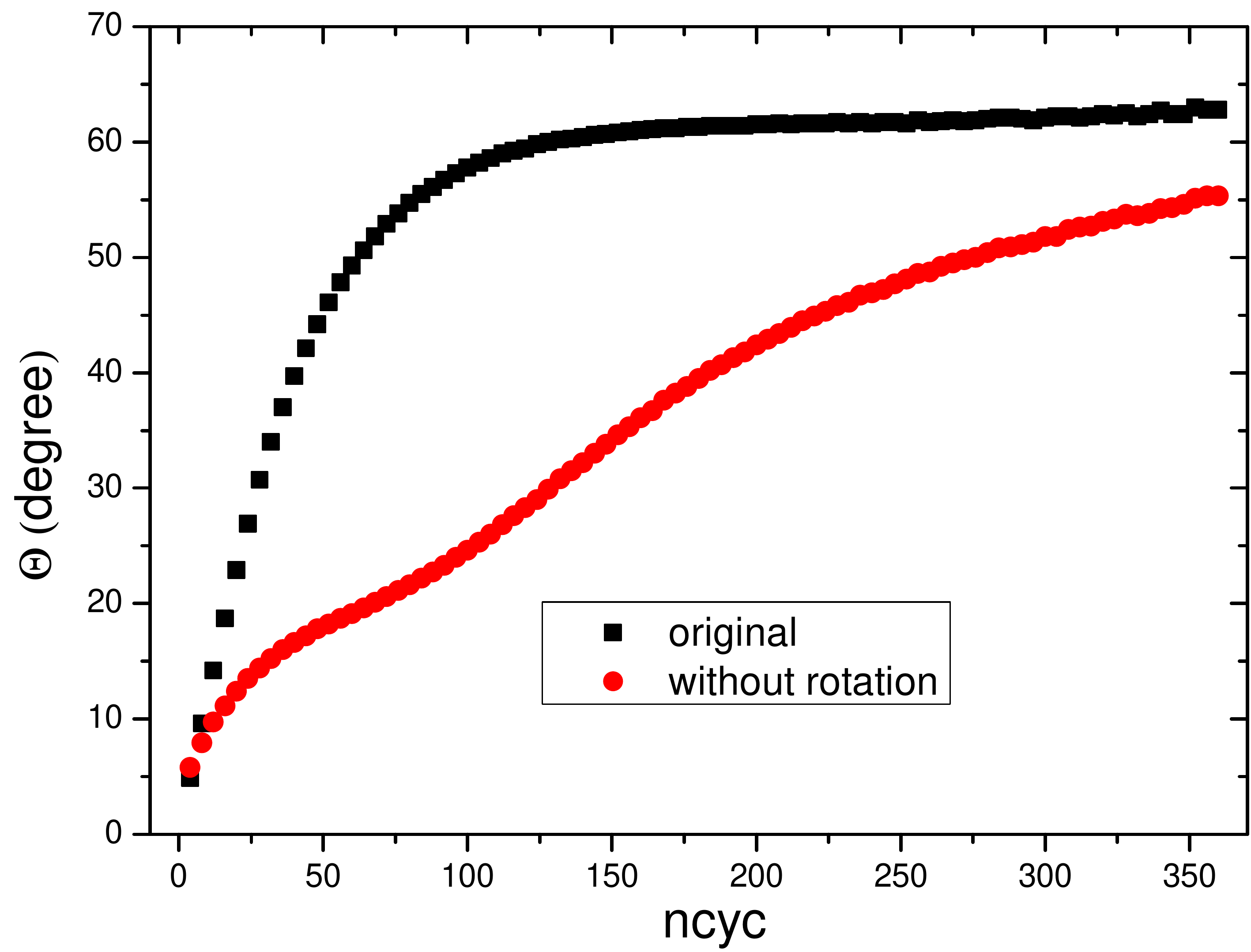}
\caption{
The change of the flow angle with time 
in a  Pb+Pb reaction at $\sqrt{s_{NN}}=2.76$TeV and $b= 0.7\, b_{max}$,
for the
flow tensor for the original flow evolution (black points), while 
the flow angle development of the flow tensor where the rotational
component of the velocity was removed is also shown (red dots).
The DCF was evaluated and presented at the time-step, $ncyc = 84$
(corresponding to $t = 3.56$fm/c),
where the flow angle of the rotation-less flow tensor is 22.2 degrees.
From \cite{CsW13Koly}.
}
\label{fig-9}       
\end{figure}

On the other hand this is not true if the local velocities have
a "side" component, i.e. when the source is rotating. 
This can be tested by the introduction of the Differential 
Correlation function, $\Delta C(k,q)$, which is defined as
\be
\Delta C(k,q) \equiv C(k_+,q_{out}) - C(k_-,q_{out}) .
\ee
Now let us assume that the rotation axis is the $y$-axis,
the momentum vector $k$, points into the $x$ direction,
and the tilted vectors are $k_{+x} = k_{-x}$ and  $k_{+z} = - k_{-z}$.
In a heavy ion reaction $z$ could be the beam direction and the 
$x, z$ plane is the reaction plane. E.g. for central collisions
or spherical expansion, $\Delta C(k,q)$ would vanish!
It would become finite if the rotation introduces an asymmetry.

We have studied the differential $\Delta C(k,q)$ -function, and for 
symmetric sources its amplitude is increasing with the speed of 
rotation \cite{CV13}, as expected.

For realistic heavy ion collision studies we used the
same PICR fluid dynamical code as in the previous examples. The
Global Collective flow shows the same symmetry features, as described
above. At peripheral collisions the shape of the emitting
source can be approximated with a three axis ellipsoid,
and different methods exist to characterize the shape and the 
directions of the axes.  A traditional method is the Global Flow Tensor
analysis, which dates back to the 1980s. The main tilt axis of the 
emission is different with, and without rotation, See Fig. 
\ref{fig-9}.

\begin{figure}[h] 
\centering
\includegraphics[width=9cm,clip]{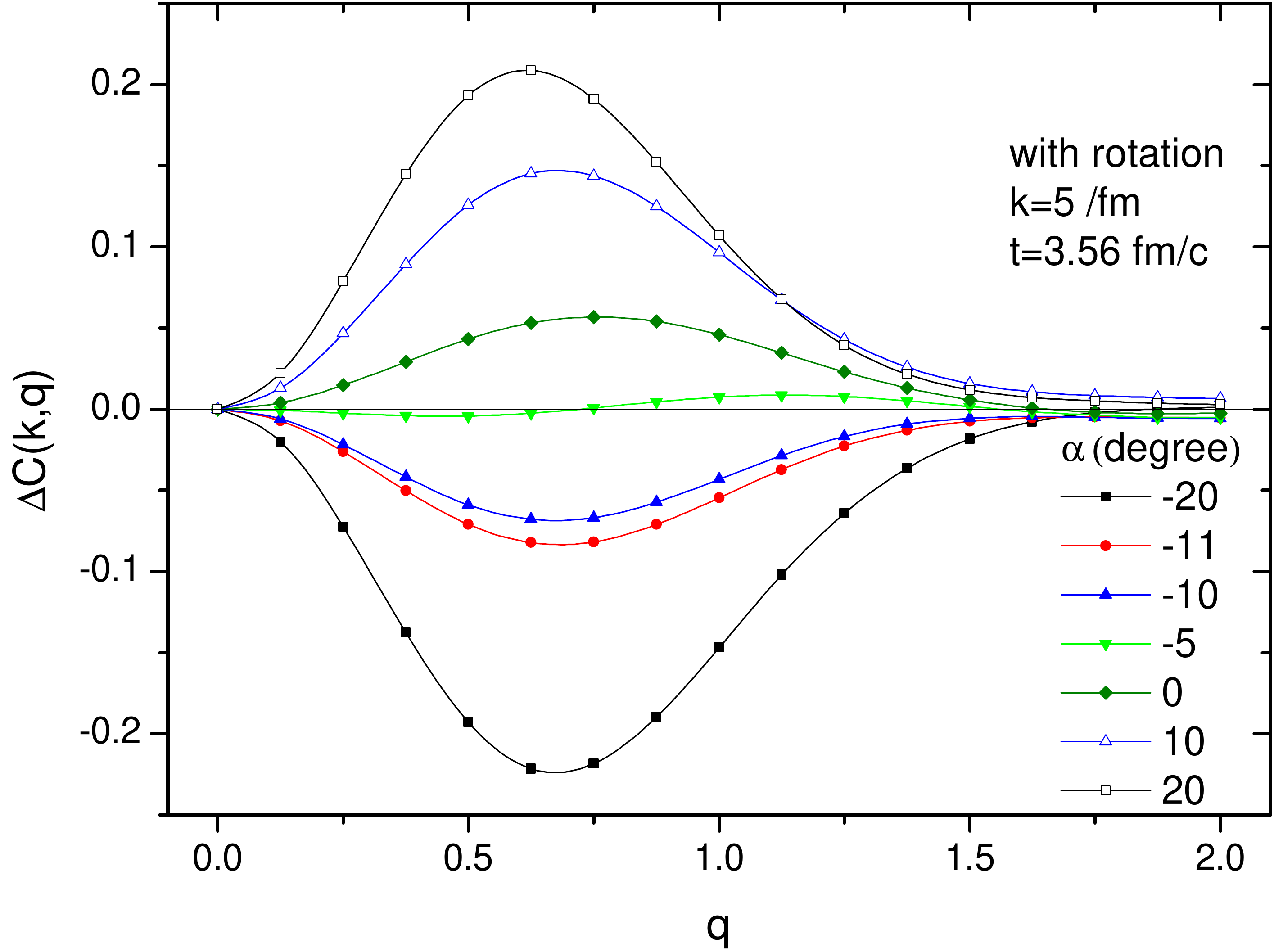}
\caption{
The Differential Correlation Function (DCF) at average pion wavenumber,
$k = 5/$fm and fluid dynamical evolution time, $t=3.56$fm/c, as a function
of the functions of momentum difference in the "out" direction $q$
(in units of 1/fm). The DCF is evaluated in a frame rotated in the
reaction plane, in the CM system by angle $\alpha$. The figure 
shows the result where the rotation component of the velocity
field is not removed with a minimum at $\alpha = -5$ degrees . 
In contrast the DCF shows a minimum in its integrated value
over $q$, for $\alpha = -11$ degrees when the rotation component of the
velocity is removed \cite{CsW13Koly}.   The shape of the DCF changes 
characteristically with the angle $\alpha$. 
The "rotation-less" configuration is constructed in the
fluid dynamical model, where the $\alpha = -11$ degree symmetry angle is 
found. Unfortunately this is not possible experimentally, so the direction
of the symmetry axes should be found with other methods, like global flow 
analysis and/or azimuthal HBT analysis.
}
\label{fig-10}       
\end{figure}

The correlation function for the original fluid dynamical
DCF with the rotation included (Fig. \ref{fig-10}) is different from the one
obtained from the rotation-less configuration (see Fig. 4 of 
ref. \cite{CVW13}. 
At  the reference frame angle corresponding
to the symmetry angle of the rotation-less system,
$\alpha = -11$ degrees, the correlation
function is distinctly different from the rotation-less one 
and has a minimum of $-0.085$ 
at $q=0.63/$fm \cite{CVW13}.
Unfortunately this is not possible experimentally, so the direction
of the symmetry axes should be found with other methods, like global flow 
analysis and/or azimuthal HBT analysis.

To study the dependence on the angular momentum the same study was 
for lower angular momentum also, i.e. for a lower (RHIC) energy Au+Au
collisions at the same impact parameter and time. We identified the 
angle where the rotation-less DCF 
was minimal, which was 
$\alpha = -8$ degrees, less than the deflection at higher angular momentum.
The original, rotating configuration 
was then analyzed at this
deflection angle, and a minimum of $-0.046$ appears at $q=0.76/$fm.
Thus, the magnitude of the DCF at the angle of the symmetry axis
increased by nearly a factor of two.

Thus, the method is straightforward for symmetric emission objects,
while for a general Global Collective Flow pattern one has to
extract the shape symmetry axis with other methods. There are 
several methods for this task, and it takes some experimental
tests, which of these methods are the most adequate for the task.

\section{Conclusions }
\label{sec-6}
%

Following the introduction of early flow studies the importance of 
splitting flow fluctuations from Global Collective Flow was discussed 
and this separation was presented.
Such a separation will enable the separate 
study the Global Collective flow component,
which includes novel new features not studied up to 
now, including rotation, turbulence, and Kelvin Helmholtz 
Instability.
As the measurement of directed flow is difficult due to its
decreasing amplitude at increasing beam energies, alternative
detection methods are presented, which are more sensitive
to these processes.

We are looking forward that these new phenomena with the
help of the suggested methods will open new ways of studying the
Quark-gluon Plasma. Especially the transport
properties are the key features, as some of the new phenomena,
like turbulence and the Kelvin Helmholtz Instability occur
only in case of low viscosity.

\section*{Acknowledgements}
Comments from Peter Braun-M\"unzinger, Joseph I. Kapusta,
Volodymyr Magas and Dujuan Wang are gratefully acknowledged.

\section*{References}



\begin{thebibliography}{}

\bibitem{g1}
Scheid W, M{\"u}ller H and Greiner W 1974
{\it Phys. Rev. Lett.} {\bf 32} 741 \\
Scheid W, Ligensa R and Greiner W 1968
{\it Phys. Rev. Lett.} {\bf 21} 1479 \\
Scheid W and Greiner W 1969
{\it Z. Phys.} {\bf 226} 364

\bibitem{g2}
Baumgardt H {\it et al} 1975
{\it Z. Phys.} A {\bf  273}  359

\bibitem{g3}
Chapline G F,  Johnson M H, Teller E and  Weiss M S 1973
{\it Phys. Rev.} D {\bf 8} 135

\bibitem{g4}
Gustafsson H \AA\ , Gutbrod H H {\it et al} 1984 
{\it  Phys. Rev. Lett. } {\bf 52} 1590

\bibitem{Cs94}
Csernai L P 1994 {\it Introdiction to Relativistic Heavy Ion Collisions}
John Wiley \& Sons, Chichester

\bibitem{cr99}                                          
Csernai L P and R{\"o}hrich D 1999
{\it Phys. Lett.} B {\bf 458} 454

\bibitem{antiflow}
Brachmann J, Soff S, Dumitru A, S\"ocker H, Maruhn J~A, 
Greiner W, Bravina L~V and Rischke D~H  2000 
{\it Phys. Rev. C} {\bf 61}, 024909

\bibitem{SnellingsEA-00-08}
Snellings R J M, Sorge H, Voloshin A A, Wang F Q and Xu N 2000
{\it Phys. Rev. Lett.} {\bf 84} 2803;
arXiv: nucl-ex/9908001

\bibitem{csg10.}  
Csernai L P and  Barz H W 1980
{\it Z. Phys.} A {\bf 296} 173

\bibitem{csg13.}
Buchwald G, Csernai L P, Maruhn J, Greiner W and St{\"o}cker H 1981
{\it Phys. Rev.} C {\bf 24} 135

\bibitem{csg23.}
Csernai L P, Lovas I, Maruhn J, Rosenhauer A, Zim{\'a}nyi J, Greiner W 1982
{\it Phys. Rev.} C {\bf 26} 149

\bibitem{csg14.}                                         
Csernai L P and  Greiner W 1981
{\it Phys. Lett.} B {\bf 99} 85

\bibitem{csg20.}  
St{\"o}cker H, Csernai L P, Graebner G, Buchwald G,
Kruse H, Cusson R Y, Maruhn J and Greiner W 1982
{\it Phys. Rev.} C {\bf 25} 1873

\bibitem{csg21.}  
Csernai L P, Greiner W, St{\"o}cker H, Tanihata I, 
Nagamiya S, Knoll J 1982
{\it Phys. Rev.} C {\bf 25} 2482

\bibitem{csg24.}
Barz H W, Csernai L P and Greiner W 1982
{\it Phys. Rev.} C {\bf 26}  740

\bibitem{csl1}
Amsden A A {\it et al} 1977 {\it Phys. Rev.} C {\bf 15} 2059
       
\bibitem{csl2}                                         
Amsden A A {\it et al}  1975 {\it Phys. Rev. Lett.} {\bf 35} 905
    
\bibitem{csg17.}
St{\"o}cker H, Riedel C, Yariv Y, Csernai L P, Buchwald G, Graebner G,
Maruhn J, Greiner W, Frankel K, Gyulassy M, Sch{\"u}rmann B,
Westfall G, Stevenson J D, Nix J R and Strottmann D 1981
{\it Phys. Rev. Lett.} {\bf 47} 1807

\bibitem{do85}
Danielewicz P and Odyniecz G 1985
{\it Phys. Lett.} B {\bf 157}  146

\bibitem{cf86}
Csernai L P, Freier P, Mevissen J, Nguyen H and Waters L 1986
{\it Phys. Rev.} C {\bf 34} 1270.

\bibitem{bc8788}
Bonasera A and Csernai L P 1987
{\it Phys. Rev. Lett.} {\bf 59}  630; 
Bonasera A, Csernai L P and Sch{\"u}rmann B 1988
{\it Nucl. Phys.} A {\bf 476} 159

\bibitem{BzTe13}                                           
Bzdak A and Teaney D 2013
{\it Phys. Rev. C} {\bf 87}, 024906 

\bibitem{AL2013d}
Abeleev B {\it et al}, (ALICE Collaboration) 2013
{\it Phys. Rev. Lett.} {\bf 111}, 232302

\bibitem{Eyyubova}
 Csernai L P, Eyyubova G and Magas V K 2012
{\it Phys. Rev.} C {\bf 86} 024912 

\bibitem{MolnarD08}
Molnar D and Huovinen P 2008
{\it J. Phys.} G {\bf 35} 104125

\bibitem{Heinz08}
Song H C and Heinz U 2008
{\it Phys. Lett.} B {\bf 658} 279;
{\it Phys. Rev.} C {\bf 77} 064901

\bibitem{BCLS94}
Bravina L, Csernai L P, L\'evai P and Strottman D 1994
{\it Phys. Rev.} C {\bf 50} 2161

\bibitem{Csorgo94}
Cs\"org\H o T and  Csernai LP  1994
{\it Phys. Lett.} B {\bf 333} 494;
arXiv: hep-ph/9406365

\bibitem{CsMi95}
Csernai LP and Mishustin IN 1995
{\it Phys. Rev. Lett.} {\bf 74 } 5005

\bibitem{CsW13Koly} 
Csernai L P and Wang D J 2014 - invited plenary talk at the -
{\it Int. Conf.  on New Frontiers  in Physics}, ICNFP2013,
{\it (Kolymbari, Greece, 28 Aug. - 5 Sept., 2013)}
EPJ Web of Conferences, {\bf 71} 00029

\bibitem{Kovtun2005}
Kovtun P K, Son D T, Starinets A O 2005 {\it Phys. Rev. Lett.} {\bf 94}, 111601 

\bibitem{SUSY}                                                     
Buchel A, Liu J T and Starinets A O 2005
{\it Nucl. Phys.} B {\bf 707} 56

\bibitem{CKM2006}
Csernai L P,  Kapusta J I and McLerran L D 2006 {\it Phys. Rev. Lett} {\bf 97} 152303 

\bibitem{KMS12-13} 	
Kapusta J I, Mueller B and Stephanov M 2013
{\it Nucl. Phys.}  A {\bf 904-905} 499c\\
Kapusta J I, Mueller B and Stephanov M 2012
{\it Phys. Rev.}  C {\bf 85} 054906

\bibitem{INPC13}
Csernai L P 2014 - invited talk at the -
{\it (Int. Nuclear Physics Conference,
Firenze, Italy, June 2-7, 2013)}
EPJ Web of Conferences, {\bf 66} 04007 

\bibitem{MCs001}
Magas V K, Csernai L P and Strottman D D 2001
{\it Phys. Rev.} C {\bf 64} 014901

\bibitem{MCs002}                                                        
Magas V K, Csernai L P and Strottman D D 2002
{\it Nucl. Phys.} A {\bf 712} 167 

\bibitem{FW13nov} 
Floerchinger S and Wiedemann U A 2013 
{\it Phys. Rev.} C {\bf 89} 034914

\bibitem{CVW13}
Csernai L P, Velle S and Wang DJ 
2014 {\it Phys. Rev.} C {\bf 89} 034916

\bibitem{BCW2013}
Becattini F, Csernai L P and Wang D J 2013
{\it Phys. Rev.} C {\bf 88} 034905

\bibitem{RPMetal19}                                                   
Rischke DH, P\"urs\"un Y, Maruhn JA, St\"ocker H and Greiner W 1995
{\it Heavy Ion Physics} {\bf 1} 309 

\bibitem{Sorge99}
Sorge H 1999 
{\it Phys. Rev. Lett} {\bf 82} 2048

\bibitem{ZFBF01}
Zabrodin E E, Fuchs C, Bravina LV and Faessler A
2001 {\it Phys. Rev.} C {\bf 63} 034902, 
arXiv: nucl-th/0006056

\bibitem{hydro1}
Csernai L P, Magas V K, St\"ocker H and D.D. Strottman 2011
{\it Phys. Rev.} C {\bf 84}  024914

\bibitem{Jia-ATLAS-12}
Jia J et al (ATLAS Collaboration) 2012
{\it J Phys. Conf. Ser.}  {\bf 389} 012013\\ 
Jia J, Radhakrishnan S and Mohapatra S 2013
{\it J. Phys. G: Nucl. Part. Phys.} {\bf 40} 105108 

\bibitem{ALICE-PLB-12}
Aamodt K {\it et al} (ALICE Collaboration) 2012 {\it Phys. Lett.}
B {\bf 708}  249 

\bibitem{GaleEtal-12}                                         
Gale C, Jeon S, Schenke B, Tribedy P and Venugopalan R 2013
{\it Phys. Rev. Lett.} {\bf 110}  012302;
arXiv:1209.6330 [nucl-th]

\bibitem{GLVB14}                                                  
Gyulassy M, Levai P, Vitev I and Biro T 2014 
{\it Int. Conf. QM14, Darmstadt, Germany, 
May 21-25, 2014}; arXiv: 1405.7825 [hep-ph]  

\bibitem{schenke}
Schenke B, Jeon S Y and Gale C 2010 
{\it Phys. Rev.} C {\bf 82} 014903

\bibitem{Bozek}
Bozek P and Wyskiel I 2010
{\it Phys. Rev.} C {\bf 81} 054902

\bibitem{Adil}
Adil A and Gyulassy M 2005
{\it Phys. Rev.} C {\bf 72} 034907

\bibitem{Karpenko}                                               
Karpenko Iu, Huovinen P and Bleicher M 2013 arXiv: 1312.4160v1[nucl-th]

\bibitem{CK1984}
Csernai L P and Kapusta J I 1984 {\it Phys. Rev.} D {\bf 29} 2664

\bibitem{CK1985}
Csernai L P and Kapusta J I 1985 {\it Phys. Rev.} D {\bf 31} 2795
  
\bibitem{Mishu02}
Mishustin I N and Kapusta J I 2002
{\it Phys. Rev. Lett.} {\bf 88} 112501;
Mishustin I N and Lyakhov KA 2012 
{\it Phys. Atom. Nucl.} {\bf 75} 371 

\bibitem{CGC2001}
Bjoraker  J, Venugopalan R 2001 {\it Phys. Rev.} C {\bf 63}, 024609;
Iancu E, Leonidov A, McLerran L 2001 {\it Phys. Lett.} B {\bf 510}, 133;
Iancu E, Leonidov A, McLerran L 2001 {\it Nucl. Phys.} A {\bf 692}, 583.

\bibitem{GyCs86}
Gyulassy M and Csernai L P 1986
{\it Nucl. Phys.} A {\bf  460} 723    

\bibitem{IP-Sat03}
Bartels J, Golec-Biernat K J and Kowalski H 2002
{\it Phys. Rev.} D {\bf 66} 014001;
Kowalski H and Teaney D 2003
{\it Phys. Rev.} D {\bf 68} 114005

\bibitem{IP-Glasma}
Schenke B, Tribedy P and Venugopalan R 2012
{\it Phys. Rev. Lett.} {\bf 108} 252301;
{\it Phys. Rev.} C {\bf 86} 034908

\bibitem{Vov13}
Vovchenko V, Anchishkin D and Csernai L P 2013
{\it Phys. Rev.} C {\bf 88} 014901

\bibitem{hydro2}
Csernai L P, Strottman D D and Anderlik C 2012
{\it Phys. Rev.} C {\bf 85} 054901
 
\bibitem{WNC13}
Wang D J, N\'eda Z and Csernai L P 2013 
  {\it Phys. Rev.} C {\bf 87} 024908

\bibitem{McInnes}                                 
McInnes B 2014 arXiv: 1403.3258 [hep-th]

\bibitem{CMW12}
Csernai L P, Magas V K and Wang D J 2013
{\it Phys. Rev.} C {\bf 87} 034906

\bibitem{Stefan}
Floerchinger S and Wiedemann U A 2011 {\it J. High Energy Phys.} {\bf 11} 100\\
Floerchinger S and Wiedemann U A 2011 {\it J. Phys. G} {\bf 38} 124171

\bibitem{Baznat}
Baznat M, Gudima K, Sorin A and Teryaev O 2013
{\it Phys. Rev.} C {\bf 88} 061901

\bibitem{CV13}
Csernai L P and Velle S 2014 
(arXiv: 1305.0385 [nucl-th] and arXiv: 1405.7283 [nucl-th])
{\it Int. J. Mod. Phys. E  in press}

\end{thebibliography}
\end{document}